\documentclass{article} 
\usepackage{fix-cm}
\usepackage{amssymb}
\usepackage{amsmath}
\usepackage{graphicx}
\usepackage{subfigure}
\usepackage{makeidx}
\usepackage{multicol}

\usepackage{amsthm}
\newtheorem{theorem}{Theorem}[section]
\newtheorem{definition}{Definition}[section]
\newcommand*\diff{\mathop{}\!\mathrm{d}}

\begin{document}
\title{Statistics of Extremes for the Insurance Industry\footnote{to appear in Handbook of Statistics of Extremes, Chapman \& Hall}}

	\author{Hansj\"{o}rg Albrecher\thanks{Corresponding Author. Department of Actuarial Science, Faculty of
			Business and Economics,  Swiss
			Finance Institute and Expertise Center for Climate Extremes (ECCE), University of Lausanne, CH-1015 Lausanne. Email: hansjoerg.albrecher@unil.ch}, Jan Beirlant\thanks{Department of Mathematics, KU Leuven, Belgium and Department of Mathematical Statistics and Actuarial Science, University of the Free State, South Africa. Email: jan.beirlant@kuleuven.be}}
\date{}
	\maketitle
\bigskip\abstract{\begin{quote}
		\noindent 
 We provide a survey of how techniques developed for the  modelling of extremes naturally matter in insurance, and how they need to and can be adapted for the insurance applications. Topics covered include truncation, tempering, censoring and  regression techniques. The discussed techniques are illustrated on concrete data sets.
\end{quote}}	

\section{Introduction}\label{intro}\index{extreme value index}\index{extreme quantile estimation}\index{return period estimation}\index{premium estimation}
Modelling of largest claims naturally arises in actuarial practice. And insurance applications traditionally have been one of the prime applications of extreme value statistics. At the same time, the concrete challenges and specific constraints concerning model assumptions and available data in insurance practice also have been, and still are, a generator of interesting theoretical problems in extreme value statistics. \\
The aggregate sum of claim payments in an insurance portfolio is naturally dominated by the larger components, so that a thorough understanding of the tails of claim distributions is crucial for a proper pricing and management of insurance risk. Tail modelling is most important for reinsurance\index{reinsurance} companies, as the handling of extremes is (among others) one of the main responsibilities of reinsurers. \\

\begin{figure}[htb]\label{marine}
\centerline{	\includegraphics[height=90pt]{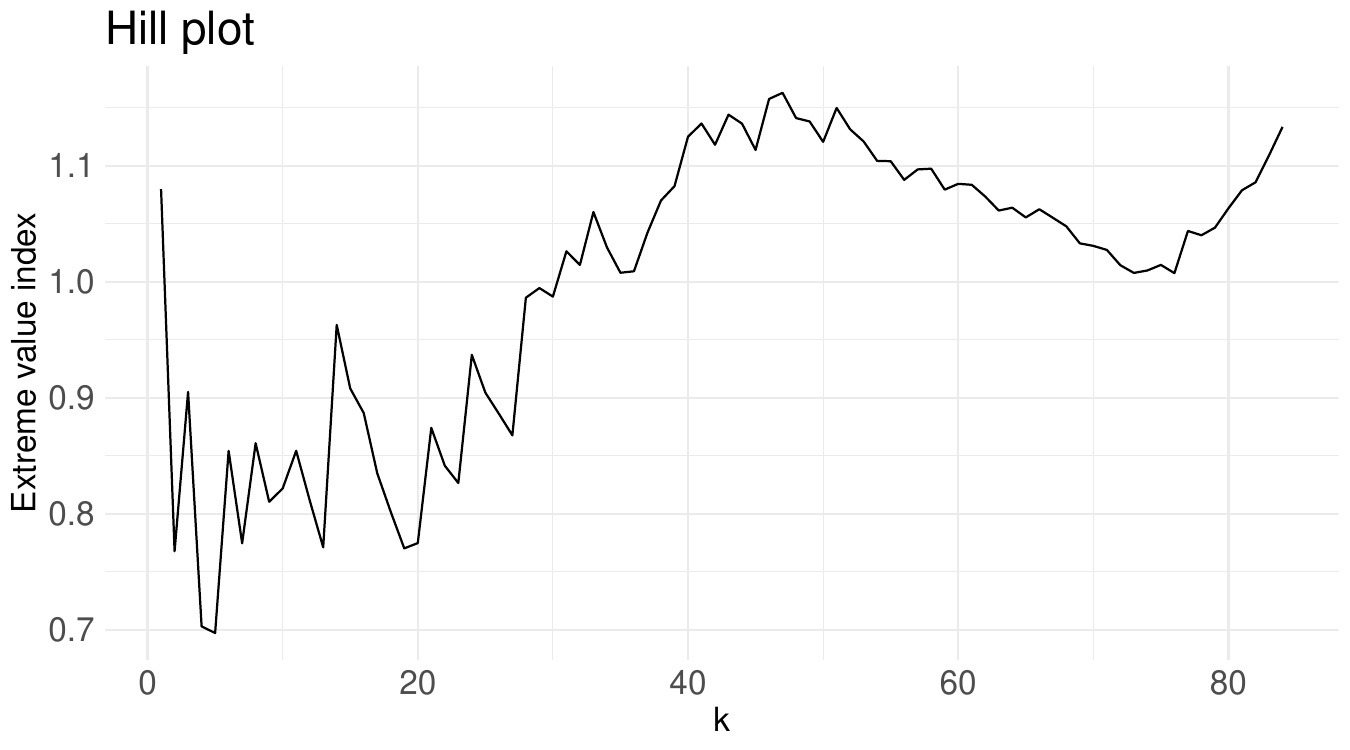}\hspace*{0.5cm}\includegraphics[height=90pt]{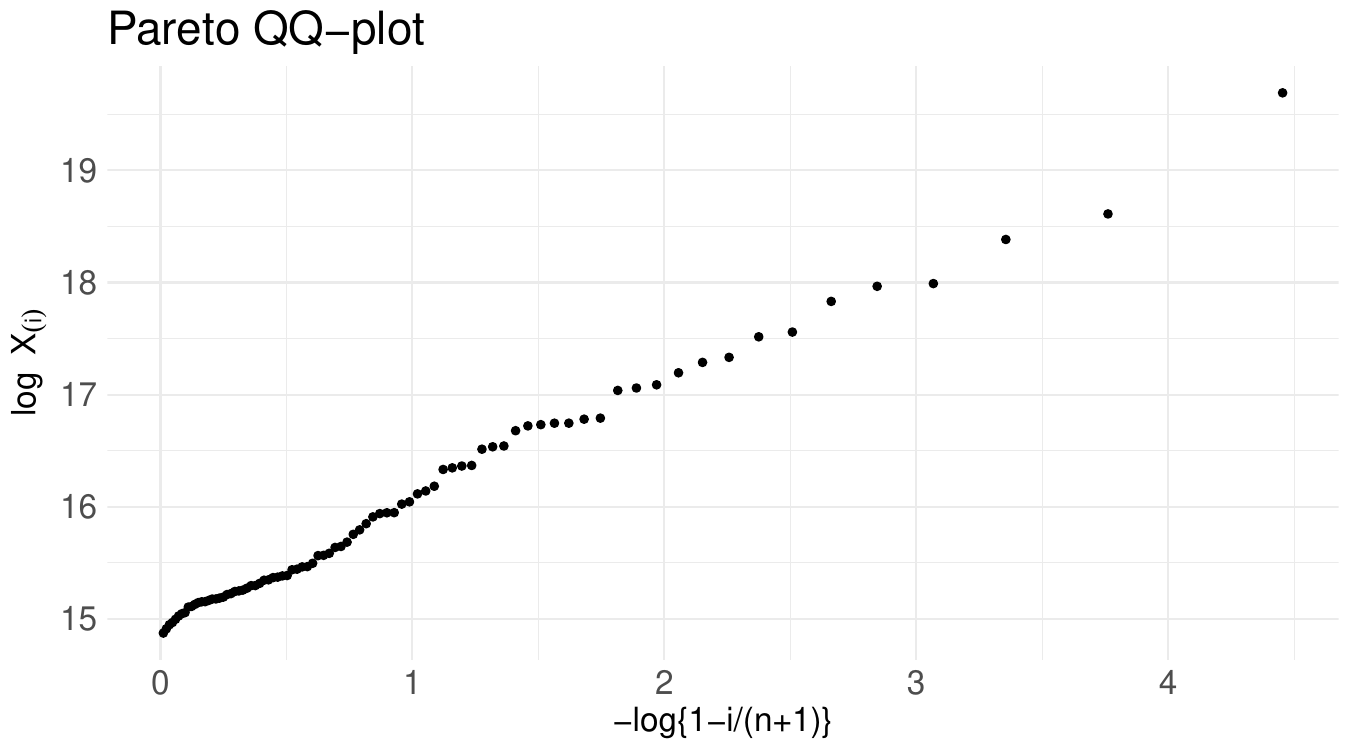}}
	\caption{Hill plot and Pareto QQ-plot of marine liability claim data, cf. Guevara-Alarc\'on\ \cite{Guevara20}.}
\end{figure}

For the appearance of very heavy tails in insurance, the marine liability data, as analyzed by Guevara-Alarc\'on et al.\ \cite{Guevara20}, can serve as an illustrative example. There, a data set of large losses from the marine line of business is statistically analyzed in a detailed manner both in terms of frequency and severity. 
Figure \ref{marine} reproduces the Hill plot\index{Hill plot} and the Pareto QQ-plot \index{Pareto QQ-plot}of these claims. One observes that the extreme value index runs between 0.8 and 1.1. This means that the existence of the first moment is questionable, while the second moment is clearly infinite. Hence, the classical risk-theoretical principles do not apply, and a simple insurance coverage is not possible.
%Catastrophic risk (CatRisk) 
 The tail risk is then typically taken over by reinsurers against a negotiated premium. 
 %Guevara-Alarc\'on et al.\ \cite{Guevara20} examine
% the popular excess-of-loss (XL) reinsurance principle. 
In general, the actuarial methods that have been developed to measure and control risk need to be tailored to each concrete situation (see, e.g.\ \cite{albrecher2010reinsurance} for a short overview and \cite{ABT17} for a detailed one). In particular, when the focus is on the tails, specific modelling situations and challenges with data availability occur, for which standard extreme value methods have to be adapted. In this chapter, we will discuss a number of such situations. %, with particular attention for risk measurement. 
Classical problems in extreme value analysis (EVA),  such as threshold selection\index{threshold selection} or bias reduction\index{bias reduction}, are mentioned where available. We also deal with the estimation of insurance premiums and risk measures based on the constructed models. \\

In Section \ref{sec12}, we introduce common reinsurance forms. There, we also present those constraints on model choices and data availability in (re)insurance practice that affect the study of extremes and lead to deviations from standard extreme value theory and statistics, in order to motivate the subsequent sections. In reinsurance settings, the Pareto-type model\index{Pareto-type distribution} often serves as a starting model, where a loss random variable $X$ is defined through
$$
P(X>x)= x^{-1/\xi}\ell (x),
$$
with $\xi>0$ the extreme value index and $\ell$ a slowly varying function defined by 
$$
\lim_{t \to \infty}\frac{\ell(xt)}{\ell(t)}=1.
$$
Section \ref{sec2} discusses adaptations of the classical tail analysis for truncated losses, as well as the less restrictive situation of tempering. In Section \ref{sec3}, we deal with the case of extremes for censored data. In insurance practice, the modelling of the body of the claim distribution and its tail is often separated (distinguishing \textit{attritional}\index{losses!attritional} and \textit{large} claims\index{losses!large}), and Section \ref{sec5} describes techniques to reconciliate the two into one model. Section \ref{sec6} and Section \ref{sec62} offer a discussion and literature overview of regression and multivariate settings, respectively, with attention for the specific modelling aspects listed above. The importance of a profound understanding of extremal risks is further exacerbated when it comes to the modelling of insurance losses due to natural catastrophes. In particular, dependence assumptions of risks across space and time as well as the non-stationarities of such risks in the light of climate change are to be considered. In Section \ref{sec7}, we finish with a respective discussion. 

\section{Reinsurance and Data}\label{sec12}
Insurance companies typically pass on their exposure to potentially very large claims to reinsurance companies. Therefore, the modelling of extremal events is mainly a concern for the latter. Reinsurers then look for means to diversify these risks, often on a global scale. In the following, we will briefly describe the most common reinsurance forms used in practice and their suitability for the protection against extremal claims. For a comprehensive treatment with attention for EVA, we refer to \cite{ABT17}.
 
\begin{itemize}
\item {\bf Proportional Reinsurance.}\index{reinsurance!proportional} For an individual risk $X_i$, a \textit{quota-share} (QS)\index{reinsurance!quota-share} reinsurance treaty is simply a proportional risk sharing with a reinsured amount $R_i=a\, X_i$ for some proportionality factor $0<a<1$. It is easy to implement for an entire portfolio, leads to a natural premium scheme (proportional premiums with adjustments for claim acquisition, handling costs etc.) and does not introduce any additional mathematical complexity, as it simply rescales the original quantities. Since this contract also reinsures small claims which the insurer could easily handle himself, a popular variant is \textit{reinsurance!surplus}\index{reinsurance!surplus}. In this case, the proportionality factor for each risk depends on the sum insured in the underlying policy, with a larger reinsured proportion for larger sum insured, and no reinsurance coverage for sums insured that are below some threshold. While these reinsurance forms are very popular in certain lines of business (e.g.\ fire and property), they do not improve the shape of the tail of the risk exposure for the insurer. This is why non-proportional coverages such as the reinsurance forms described next are much more relevant when it comes to extremes.
	
\item {\bf Excess-of-Loss (XL) Reinsurance.}\index{reinsurance!excess-of-loss}
For a given pre-defined retention $M$ in an XL treaty, the reinsurer agrees to pay for each claim in the portfolio 
the excess over the retention $M$. Typically, this will only be agreed upon up to a certain limit $L$, leading to the reinsured amount
\begin{equation}\label{tthat}
    R=\sum_{i=1}^{N}\min\{(X_{i}-M)_+,L\}
\end{equation}
for individual risks $X_i$, with $(X - u)_+ := \max(X - u, 0)$, and $N$ the number of claims in the portfolio (the insurer will then often look for additional XL treaties with higher layers until the remaining risk is considered `negligible'). When $L=\infty$, the intricate relation between an unbounded XL treaty and POT methodology modelling becomes apparent. Indeed, the modelling of  exceedances $P(X_i-M > x\mid X_i>M)= {P(X_i>M+x)}/{P(X_i>M)}$ for large values of $M$ is at the heart of EVA.\\
If one first aggregates all claims and then applies this principle, i.e.\ $R=\min\{(\sum_{i=1}^{N}X_{i}-M)_+,L\}$, then this contract is referred to as a (bounded) \textit{Stop-Loss} reinsurance treaty.\index{reinsurance!stop-loss} Particularly for the reinsurance of catastrophe losses, it is quite popular to collect all claims due to a particular event, consider them as one claim $X_i^c$, and apply an aggregate retention $M^c$ (and limit $L^c$) to get
$$
R=\sum_{i=1}^{N^c}\min\{(X^c_{i}-M^c)_+,L^c\},
$$
which is known as (bounded) \textit{cumulative XL} (or \textit{CAT-XL}).\index{reinsurance!cumulative XL} This is an attractive cover against frequency risk, namely the risk to face many claims, whose individual size may not be extreme but their sum is. Mathematically, this leads to the very same analysis, the only difference being the modelling of the distribution of $X_i^c$ and the one of the number of catastrophes $N^c$. 

\item {\bf Large Claim Reinsurance.} From the perspective of extreme value theory, it seems natural to also consider contracts of the form 
\[R=\sum_{i=1}^{r}X_{(N-i+1)}\]
for the order statistics $X_{(1)} \leq \cdots  \leq X_{(N)}$. This reinsurance form covers the $r$ largest claims in the portfolio. Variants are \textit{drop-down XL}\index{reinsurance!drop-down XL}
\[R=\sum_{i=1}^{N}\min\{L_i,(X_{(N-i+1)}-M_i)_+\},\]
where retentions and layer sizes are applied to the respective order statistics, and \textit{ECOMOR} reinsurance\index{reinsurance!ECOMOR}
\[R =  \sum_{i=1}^{N}\left(X_{i} - X_{(N-r)}\right)_+,\]
where the $(r+1)$th largest claim serves as the retention level in an otherwise standard (unbounded) XL contract. While such large claim reinsurance contracts have been implemented in practice to some extent, they have not become popular, partly because the calculation of premiums and the modelling of the retained risk is much more complicated and hence not practical. We will therefore focus on the XL type in the sequel. 
\end{itemize} 

\noindent In actuarial data sets for XL reinsurance, one faces various kinds of {\em incomplete data}:
\begin{itemize}
	\item In a bounded XL treaty, the reinsurer only pays up to a limit $L$ for each claim, and the insurer does not necessarily share the full claim amount beyond that limit with the reinsurer, and neither the information on claims that remained below the retention $M$. Hence, {\em censoring}, and in particular {\em right censoring}, at high values is to be considered. 
	\item Claims that have been incurred can still be {\em missing} due to reporting delays. Such missing data are referred to as `Incurred But Not Reported' (IBNR) data. \index{IBNR}
	\item Right-censoring also occurs when a claim has not been settled at the evaluation date, leading to `Reported But Not Settled' (RBNS) claims.\index{RBNS} In case of upper limits on the underlying policy, the real costs are censored at a fixed value. The settlement of the eventual claim amounts can take very long (even decades in catastrophe and liability insurance), and the already paid amount is a natural lower bound for the eventual claim size, which also leads to right-censoring. As reported data are then not exact, one needs to work with estimates for the final value. Here,  one distinguishes between `ultimate' estimates for non-closed claims (based on modelling assumptions, often chain ladder development methods) and `incurred' values (which are estimates from an accounting perspective, often with expert predictions on the concrete case). Such situations constitute significant challenges to an EVA.
	\item Since extreme events are rare, there are often not many data points available for an analysis. Therefore, one needs to merge actual historical data points with ones of related risks and with expert opinions (taking into account the implied uncertainty). In some cases, one even needs to price risks without a single data point (see e.g.\ Fackler\ \cite{fackler2022premium}). 
\end{itemize}

\noindent On the modelling side, there often is an upper limit of the exposure (either due to contractual limits or due to physical limits, such as the overall building values in an area for the insurance of natural catastrophe claims). This results in {\em truncation modelling}.\index{truncation} Also, claim payments are influenced by claim management and claims may, for instance, be subject to a higher level of inspection at highest damage levels leading to weaker tails than apparent from intermediate claims, leading to {\em tempering}\index{tempering} of claim distributions (cf.\ Section \ref{sectemp} for more details). Different models at different claim level intervals are not uncommon.\\

For an unlimited XL\index{reinsurance!excess-of-loss} treaty with retention $M$,  the expected reinsured amount ${E}(R_i)$ of a single claim $X_i$ with distribution function $F$ and finite mean, is given by 
\begin{eqnarray*}
\Pi(M) := {E}\{(X_i-M)_+\} = \int_M^{\infty}\{1-F(z)\}\diff z,
\end{eqnarray*}
which is also referred to as the \textit{pure premium}\index{pure premium} for $R_i$. 
Note that
\[
\Pi(M) = e(M)\overline{F}(M)
\]
with $e(u)= {E}(X_i-u\mid X_i>u )$ the mean excess function which serves as a basic tool in EVA expressing the expected POT, for which estimation procedures are available in extreme value statistics.
\\
With a finite layer size $L$ in the XL treaty, the pure premium becomes
\begin{equation*}
{E}(R_i)= \int_M^{M+L}\{1-F(z)\} \diff z = \Pi (M) - \Pi (M+L).
\end{equation*}
\noindent Considering a retention level $M$ as an upper quantile or \textit{Value-at-Risk} (VaR),\index{value-at-risk (VaR)} i.e. $\mbox{VaR}_{1-p}=Q(1-p)=\inf\{x: F(x) \geq 1-p\}$, the pure premium 
 is related to the \textit{Conditional Tail Expectation}\index{conditional tail expectation (CTE)} $\mbox{CTE}_{1-p}(X)$ (also known as expected shortfall in finance) defined by 
\begin{eqnarray}
\mbox{CTE}_{1-p}(X) &=& {E}(X\mid X>Q(1-p)) \nonumber \\
% &=& Q(1-p) + {E} \left( X- Q(1-p) | X> Q(1-p)\right)\nonumber \\
 &=& \mbox{VaR}_{1-p} (X)+ e(Q(1-p)). \nonumber 
  \label{TVAR}
 \end{eqnarray}
One immediately observes 
\[\mbox{CTE}_{1-p}(X) = \mbox{VaR}_{1-p}(X) + \frac{\Pi \{\mbox{VaR}_{1-p}(X)\}}{p}.\]
Hence, the estimation of $\mbox{VaR}_{1-p}(X)$ and  
$\Pi(M)$ at small and intermediate values of $p$, or correspondingly at high and intermediate values of $M$, is an important building block for measuring and managing risk. However, note that the commercial values of $M$ and $L$ not necessarily correspond to statistically optimal threshold values at which tail models fit well. This leads to the need of extending the tail models to {\em full models}\index{full models} fitting over larger outcome sets than the thresholds following from an EVA.  So, actuaries also need to construct models that simultaneously  fit well to the tail and to other parts of the possible outcome set. Such full models can for instance be constructed using splicing models (cf.\ Section \ref{sec5}).

\section{Adaptations of the Classical Tail Analysis}\label{sec2}
In the following we discuss two adaptations of classical EVA that are useful for insurance applications. 
\subsection{Truncation}
\begin{figure}[htb]
	\begin{center}
	    \includegraphics[height=120pt,width=120pt]{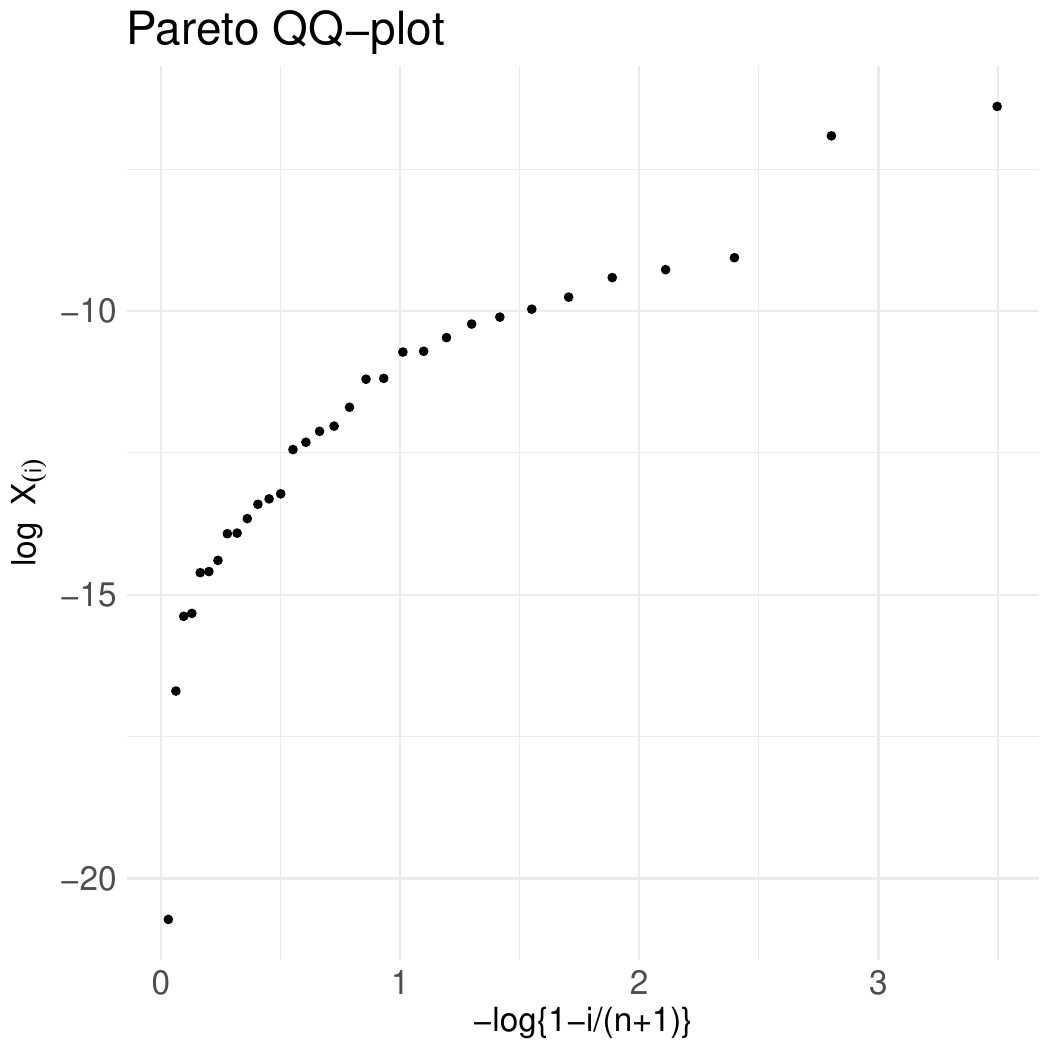}\hspace*{1cm}\includegraphics[height=120pt,width=120pt]{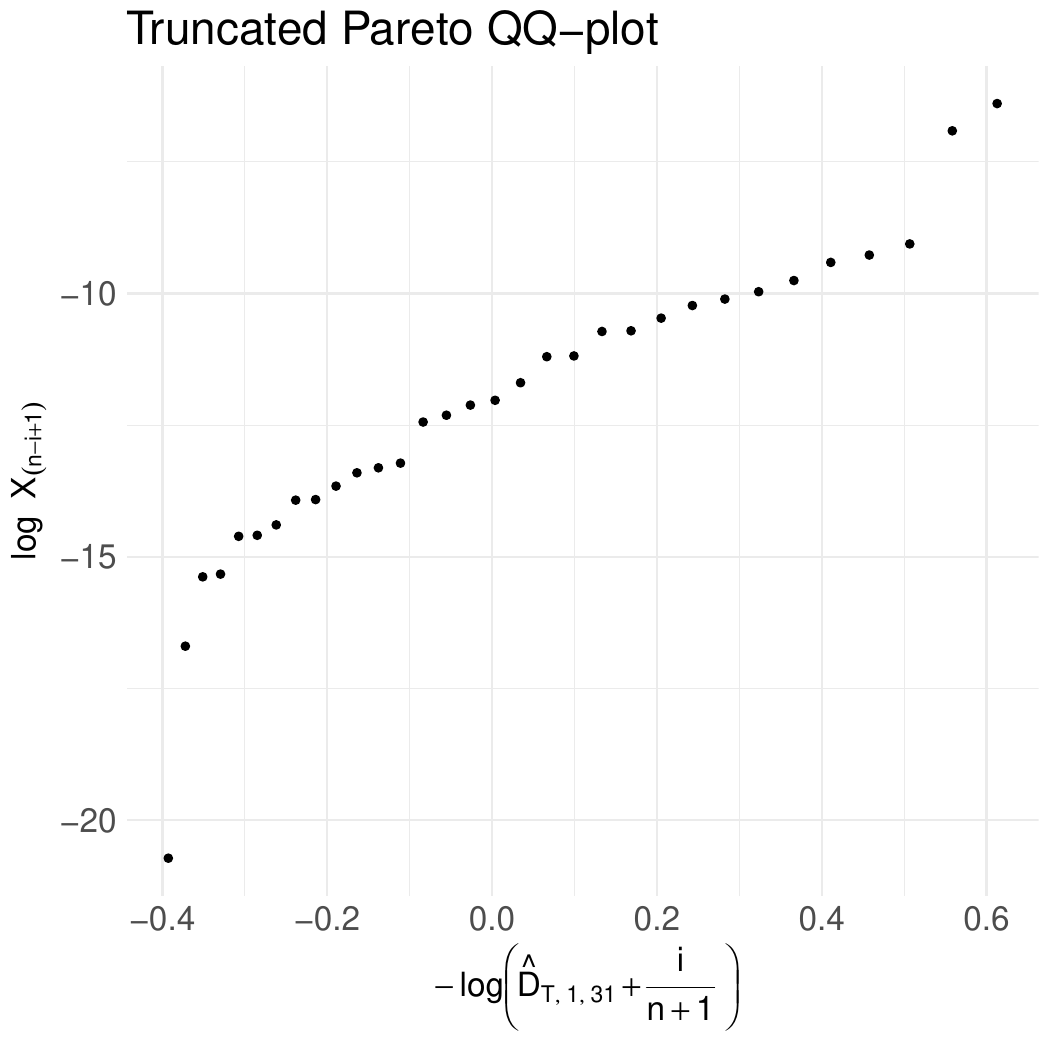}
	\caption{Pareto QQ-plot (left) and truncated Pareto QQ-plot (right) for German flood loss data.}\label{DEfloods}
	\end{center}
\end{figure}

Practical problems can arise when using the strict Pareto distribution and the more general Pareto-type model because some probability mass can still be assigned to loss amounts that are unreasonably large or even impossible. With respect to tail fitting of insurance claim data, 
upper-truncation is of interest and can for instance be due to the existence of a maximum possible loss due to policy or treaty limits or physical limits (like building values in catastrophe insurance). Such truncation effects are sometimes visible in data, for instance when an overall linear Pareto QQ-plot  shows non-linear deviations at only a few top data (see Figure \ref{DEfloods} (left)). 
%Of course such effects can also happen for log-normal or other distributions in the Gumbel domain. 
Let $W$ be an underlying non-truncated distribution with distribution function $F_W$ and quantile function $Q_W$. Upper-truncation of the distribution of $W$ at some deterministic value $T$ then leads to the truncated loss
$W\mid W<T$, whose distribution function we denote by $F_T$. In practice one does not always know if the data $X_1,\ldots,X_n$ come from a truncated or non-truncated distribution. As a consequence, the behaviour of estimators should be evaluated under both cases, and a statistical test for upper-truncation is useful. This section is taken from Beirlant et al.\ \cite{BFG16}, where the Pareto-type case is treated. Aban et al.\ \cite{aban2006parameter} considered the strict Pareto case. In Beirlant et al.\ \cite{beirlant2017fitting}, the case where $W$ belongs to any max-domain of attraction is considered. 
\\

Upper-truncation\index{truncation} of $W$, at some truncation point $T$, yields 
$$
\overline{F}_T (x) = \frac{P(W>x) - P(W>T)}{1- P(W>T)} =\frac{\overline{F}_W(x) - \overline{F}_W(T)}{F_W(T)},
$$
and the corresponding quantile function $Q_T$ is given by
$$
Q_T(1-p) = Q_W [1- \{\overline{F}_W(T) +p (1-\overline{F}_W(T)) \}] = Q_W [\{1-\overline{F}_W(T)\}(1-p)]
$$
%while the tail function $U_T$ satisfies
%\begin{equation}
%U_T(y) = U_W \left(1/\{\overline{F}_W(T) +{F_W(T) \over y}   \}\right) = U_W \left({1 \over \overline{F}_W(T) }[1+{F_W(T) \over y \overline{F}_W(T) }]^{-1}\right),
%\label{yDTaway0}
%\end{equation}
or 
\begin{equation}
Q_T(1-p) 
= Q_W \left\{ 1-\overline{F}_W (T)\left(1+ {p \over D_T}\right) \right\},
\label{yDT0}
\end{equation}
with  $ D_T := \overline{F}_W(T)/ F_W(T) $ being the odds of the truncated probability mass under the non-truncated distribution $W$. Note that for a fixed $T$,  upper-truncation models exhibit an extreme value index $\xi=-1$. For instance, in case of a simple Pareto distribution one has
\begin{eqnarray*}
Q_T(1-p) &=& \left( p (1-T^{-1/\xi}) + T^{-1/\xi}\right)^{-\xi} \\
&=& T\left(1+p \frac{1-T^{-1/\xi}}{T^{-1/\xi}} \right)^{-\xi} \\
&=& T\left(1-\xi p \frac{1-T^{-1/\xi}}{T^{-1/\xi}}(1+ o(1)) \right) \mbox{ as } p \to 0.
\end{eqnarray*}
%This follows from verifying \eqref{Cgamma} for $U_T$ as given in \eqref{yDT0}. For instance when $U_W (x) = x^{\xi}$, we find $(\overline{F}_W (T))^{\xi}U_T(x)= (1+{1 \over xD_T})^{-\xi}= (1-{\xi \over xD_T}(1+o(1)))$ as $x \to \infty$. This final expression satisfies \eqref{Cgamma} with $\xi= -1$.
%{\bf Truncation} Beirlant et al.\ \cite{beirlant2016tail}\\

In case of a Pareto-type $W$, using POT modelling with $T/t \to \beta >0$ as $t \to \infty$, the distribution of a truncated loss $X$ leads to
\begin{eqnarray*}
{P}(X/t >y\mid X>t) &=& {P}(W/t >y \mid t<W<T) \\ 
&= & \frac{\overline{F}_W (yt)-\overline{F}_W (T) }{\overline{F}_W (t)-\overline{F}_W (T)}   \\
&=&  \frac{y^{-1/\xi}{\ell_F (yt) \over \ell_F(t)} -\left({T \over t}\right)^{-1/\xi}{\ell_F (T) \over \ell_F(t)} }{1-\left({T \over t}\right)^{-1/\xi}{\ell_F (T) \over \ell_F(t)}} \\
&\to & \frac{y^{-1/\xi}-\beta^{-1/\xi}}{1- \beta^{-1/\xi} }, \quad  1<y<\beta,
\end{eqnarray*}
with $\ell_F$  a slowly varying function. Thanks to this result, the following POT approximation will be used, assuming $T/t \to \beta >0$ as $t \to \infty$:
\begin{equation}\label{that}
    {P}(X/t >y\mid X>t)\approx \frac{y^{-1/\xi}-\beta^{-1/\xi}}{1- \beta^{-1/\xi} }, \quad 1<y<\beta.
\end{equation}
Using a high order statistic $X_{(n-k)}$ as the threshold $t$, denoting the relative excesses by $R_{j,k} := X_{(n-j+1)}/X_{(n-k)}$, $j = 1, \ldots , k$, and estimating $\beta$ by $R_{1,k}$,
the pseudo-maximum log-likelihood function is given by (upon differentiating the right-hand side of \eqref{that} w.r.t. $y$)
\begin{eqnarray*}
\log L_{k,n} (\xi ) &=& \log \prod_{j=1}^k \frac{ R_{j,k}^{-1-1/\xi }}{\xi ( 1-R_{1,k}^{-1/\xi})} \\
 &=& -k\log \xi -\left(1+{1 \over \xi}\right) \sum_{j=1}^k \log R_{j,k} -k\log\left( 1- R_{1,k}^{-1/\xi} \right),
 \end{eqnarray*}
%Now 
%\begin{equation*}
%{\partial \log  L_{k,n} (\xi ) \over \partial \xi}
%=
%-{k \over \xi} - \sum_{j=1}^k \log {X_{n-j+1,n} \over %X_{n-k,n}} 
%- k \frac{ ( {X_{n,n} \over X_{n-k,n}})^{-1/\xi} \log  %{X_{n,n} \over X_{n-k,n}} }{ 1- ( {X_{n,n} \over X_{n-%k,n}})^{-1/\xi}},
%\end{equation*}
which leads to the defining equation for the likelihood estimator $\hat{\xi}^T_{k,n}$:
\begin{equation*}
H_{k,n} = \hat{\xi}^T_{k,n} - {R_{1,k}^{-1/\hat{\xi}^T_{k,n}}\log R_{1,k} \over 1- R_{1,k}^{-1/\hat{\xi}^T_{k,n}}}.
\end{equation*}
Using \eqref{yDT0} and considering the ratio of quantiles $Q_T(1-p)$ with $p={1/(n+1)}, {(k+1)/(n+1)}$, estimating $Q_T(1-p)$ by the empirical quantiles $X_{(n)}$ and $X_{(n-k)}$ respectively, and taking $Q_W(1-p)$ as $p^{-\xi}$,  we arrive at a simple estimator of $D_T={{P}(W>T)}/{{P}(W \leq T)}$:
$$
\hat{D}_T = {k \over n} \frac{R_{1,k}^{-1/\hat{\xi}_k^T} - {1 \over k}}{1- R_{1,k}^{-1/\hat{\xi}_k^T}}.
$$
In practice, one makes use of the admissible estimator $
\widehat{D}^{(0)}_T :=  \max \{ \hat{D}_T,0\}
$
to make it useful in case of truncated and non-truncated Pareto-type distributions.\\

If $D_T >0$, we find  from \eqref{yDT0} for $W$ being strict Pareto that
\begin{equation}
Q_T(1-p) = {P}(W>T)^{-1/\xi} \left[ 1+ {p \over D_T}\right]^{-1/\xi}.
\label{simpleTpar}
\end{equation} 
 Then, in order to construct estimators of $T$ and of extreme quantiles $q_p=Q_T(1-p)$, with \eqref{simpleTpar} we find that
\begin{equation*}
\left\{ {Q_T(1-p) \over Q_T(1-{k+1 \over n+1})} \right\}^{1/\xi} = { 1+ {k+1 \over (n+1)D_T} \over  1+ {p \over D_T} } =  { D_T+ {k+1 \over n+1} \over  D_T+ p  }.
\label{ratio2}
\end{equation*}
Taking logarithms on both sides and estimating  $Q_T(1-(k+1)/(n+1))$ by $X_{n-k,n}$, we find an estimator $\hat{Q}^T_{k}(1-p)$ of $Q(1-p)$:
\begin{equation*}
\log \hat{Q}^T_{k}(1-p) = \log X_{(n-k)} + \hat{\xi}^T_{k,n}
\log \left( {\hat{D}_T+ {k+1 \over n+1} \over \hat{D}_T + p} \right),
\label{Qest1} 
\end{equation*}
which equals the Weissman estimator \cite{weissman1978} when $\hat{D}_T=0$.
An estimator $\hat{T}_{k,n}$ of $T$ follows from letting $p \to 0$ in  the above expressions for $\hat{Q}^T_{p,k}$:
\begin{equation*}
\log \hat{T}_{k,n} =  \max \left[\log X_{(n-k)} +  
\hat{\xi}^T_{k,n}\log \left\{ 1+ {k+1 \over (n+1)\hat{D}_{T}}  \right\}, \log X_{(n)}  \right].
\label{Test}
\end{equation*}
Finally, a test for a truncation setting is to consider an unbounded Pareto distribution as the null hypothesis ($H_0: \, T=\infty$) and to reject $H_0$ for small values of $p$-value given by
$$\Phi \left(\sqrt{12 k}  \frac{\overline{R}_k - 0.5}{1-\overline{R}_k} \right),
$$
where $\overline{R}_k = {1 \over k} \sum_{j=1}^k R_{j,k}^{-1/H_{k,n}}$.\\

\noindent{\bf Example: Flood Risk Pooling in Europe.} Prettenthaler et al.\ \cite{prettenthaler2017flood}  discussed some challenges in insuring flood risk in Europe and analyzed (insured) flood loss data
across Europe (provided by MunichRe NatCatSERVICE), transformed into
losses expressed as a percentage of building stock value. Here we present results for the 
aggregate annual losses for the period 1980–2013 for Germany. While the Pareto QQ-plot starts off linearly,  truncation starts to set in at the top 15 data. Based on \eqref{simpleTpar}, the truncated Pareto QQ-plot  $$\left\{\left(-\log \left(\hat{D}_T+ {j \over n+1}\right),\log X_{(n-j+1)}\right)\right\}_{j=1}^n$$
is  expected to be linear at the top data under truncation. Here the endpoint is estimated at $\hat{T}=0.28\%$. 

\subsection{Tempering}\label{sectemp}
An alternative way to incorporate that the power-law behaviour does not extend indefinitely is assuming  tapering effects. In that case, the distribution tail eventually decays more quickly than according to the power-law, representing a kind of interpolation between the truncated and non-truncated case. Inspired by applications in geophysics and finance, Meerschaert et al.\ \cite{Meerschaert2012parameter} discussed parameter estimation under exponential tempering of a simple Pareto law  with survival function
\begin{eqnarray}
{P}(X > x) = c x^{-\alpha}e^{-\beta x},
\label{expPa}
\end{eqnarray} 
where  $\alpha, \beta >0$ and $c>0$ is a scale parameter.\footnote{Note here that for such tempered distributions the extreme value index equals 0 and not $1/\alpha$, and hence we prefer not to use the notation $1/\xi$ for $\alpha$ in this subsection in order to avoid confusion.}
 In the context of insurance data, Raschke \cite{raschke2020alternative} recently discussed the use of the more general Weibull tempering of a simple power law with survival function
\begin{eqnarray}
{P}(X > x) = c x^{-\alpha}e^{-(\beta x)^{\tau}},
\label{weiPa}
\end{eqnarray}
with $c, \alpha,\beta, \tau >0$. The generalization to the Pareto-type case \index{Pareto-type distribution}was provided in Albrecher et al.\ \cite{AAB_Astin21}. \

Assume $X = \min(W, Y)$ with $W$ and $Y$ independent, where $W$ is Pareto-type distributed and
\begin{eqnarray*}
{P}(Y > x) = e^{-(\beta x)^\tau}, \quad x>0.
\end{eqnarray*}
Then for some slowly varying function $\ell$
 \begin{eqnarray*}
{P}(X > x)  = x^{-\alpha}\ell(x) e^{-(\beta x)^\tau}.
\end{eqnarray*}
\noindent Using POT modelling with $t \beta_t \to \beta_{\infty}>0$
for $x>1$ as $t\to\infty$ leads to
\begin{align*}
 {P} \left({X/t}>x \mid X>t\right)
&= \frac{(tx)^{-\alpha}}{t^{-\alpha}}\frac{\ell(xt)}{\ell(t)} \frac{e^{-(\beta_t xt)^\tau}}{e^{-(\beta_t t)^\tau}} \\
&= x^{-\alpha} \frac{\ell(xt)}{\ell(t)} e^{-(\beta_t t)^\tau (x^\tau-1)} \\
& \to x^{-\alpha} e^{-\beta_{\infty}^\tau (x^\tau-1)}.
\end{align*}
Hence, the following POT approximation \index{POT approximation}will be used, assuming 
 $t \beta_t \to \beta_{\infty}>0$ and $\lambda =\beta_\infty^\tau$:
$$
{P}(X/t >y\mid X>t)\approx x^{-\alpha} e^{-\lambda (x^\tau-1)}.
$$
Maximum likelihood estimators\index{maximum likelihood estimation} for $\xi, \lambda,\tau$ using POTs $R_{j,k}=X_{(n-j+1)}/X_{(n-k)}$, $j=1,\ldots,k$  are then obtained from
\begin{eqnarray*}
\hspace{-1cm} \log L(\alpha, \lambda,\tau) &=& -\left(1+\alpha\right)\sum_{j=1}^{k} \log R_{j,k} -\lambda \sum_{j=1}^{k} \left( R_{j,k}^\tau -1\right)\\
&& + \sum_{j=1}^{k} \log\left(\alpha + \lambda\tau R_{j,k}^\tau \right). 
\end{eqnarray*}
\noindent Another possible approach focuses on the goodness-of-fit of the tempering model to the POT data above the different thresholds $X_{(n-k)}$,
 using a QQ-plot approach. Then, for a given value of $\tau$, one finds the least-squares line through the points
\begin{equation}\label{eqline}
\left\{\left(-\log \{1- \hat{F}_k(R_{j,k})\},  \alpha \log R_{j,k} + \tau \beta_{\infty}^{\tau} h_{\tau}( R_{j,k})\right)\right\}_{j=1}^k,
\end{equation}
where $h_\tau (x) = (x^\tau -1)/\tau$ and $\hat{F}_k$ denotes the empirical distribution function based on  $R_{j,k}$, for $j=1,\ldots,k$. Therefore, since $1- \hat{F}_k\left(R_{j,k}\right)= (k-j+1)/(k+1)$, one is led to minimize the weighted least squares
\begin{align*}
\mbox{WLS}(R_{j,k}; \alpha_k, \delta_k, \tau_k) := \sum_{j=1}^k w_{j,k} \left\{ \frac{1}{\alpha} \log \frac{k+1}{k-j+1} - \log R_{j,k} - \delta h_{\tau}\left( R_{j,k}\right) \right\}^2
%\label{WtempPar_QQ}
\end{align*}
with respect to $\alpha$ and $\delta=\tau \beta_{\infty}^{\tau}/\alpha$, where $\{w_{j,k}, j = 1, \ldots, k\}$ are appropriate weights. In particular, if $w_{j, k} = 1/\log \{ (k+1)/(k-j+1) \}$ when $\delta \downarrow 0$, i.e.\ without tempering, we recover the classical Hill estimator $H_{k,n}$. Optimization of WLS  also leads to an adaptive selection method for choosing $k$ which gives  appropriate estimates for $(\alpha, \tau, \beta_{\infty})$, choosing the $k$ for which the WLS value is minimal.\\

To estimate return periods of the type $1/{P} (X>c)$ for some large outcome level $c$, one can use the approximation
\[
\frac{{P}(X>tx)}{{P} (X>t)}\approx x^{-\alpha}e^{-\lambda \tau h_{\tau}(x)}
\]
with $t$ large. Setting $tx=c$ and $t=X_{(n-k)}$ for some $k$ yields an estimator for ${P} (X>c)$:
\begin{equation*}
\hat{P}_{c,k} = {k+1 \over n+1}\left( {c \over X_{(n-k)}}\right)^{-\hat{\alpha}_k }\exp \left\{-\hat{\lambda}_k \hat{\tau}_k h_{\hat{\tau}_k} \left({c \over X_{(n-k)}}\right)\right\},
\label{prob}
\end{equation*}
where $\hat{\alpha}_k, \hat{\tau}_k, \hat{\lambda}_k $ denote the estimators of these parameters based on the yop $k$ observations. 
The value $c=\hat{Q}^W_{p,k}$ solving the equation 
\begin{equation*}
{k+1 \over n+1}\left( { c \over X_{(n-k)}}\right)^{-\hat{\alpha}_k }\exp \left\{-\hat{\lambda}_k \hat{\tau}_k h_{\hat{\tau}_k} \left({c \over X_{(n-k)}}\right)\right\}=p,
\label{quant}
\end{equation*}
for a given value $p \leq {1 /n}$ then is an estimator for an extreme quantile or return level $Q(1-p)=\text{VaR}_{1-p}$ of $X$.\\

\begin{figure}[htb]
\centerline{\includegraphics[height=90pt]{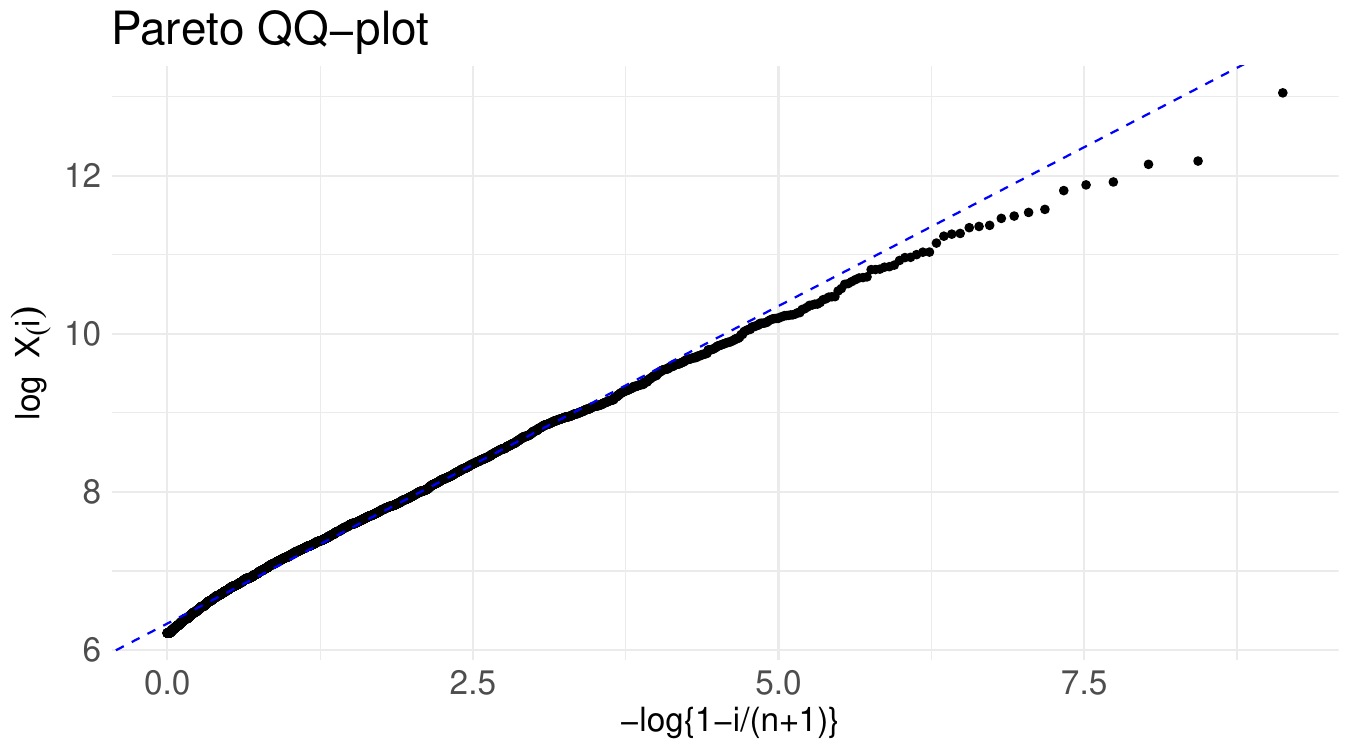}\hspace*{0.5cm}\includegraphics[height=90pt]{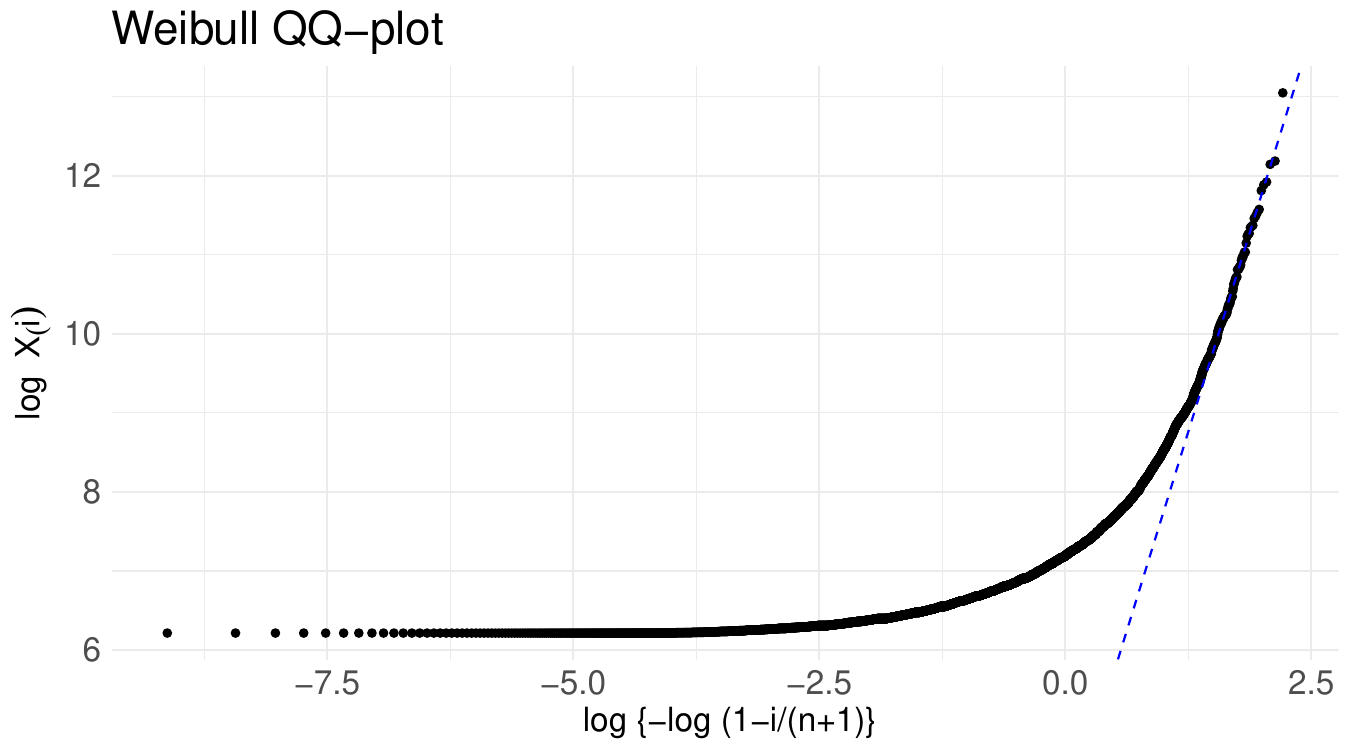}}

\centerline{
	\includegraphics[height=90pt]{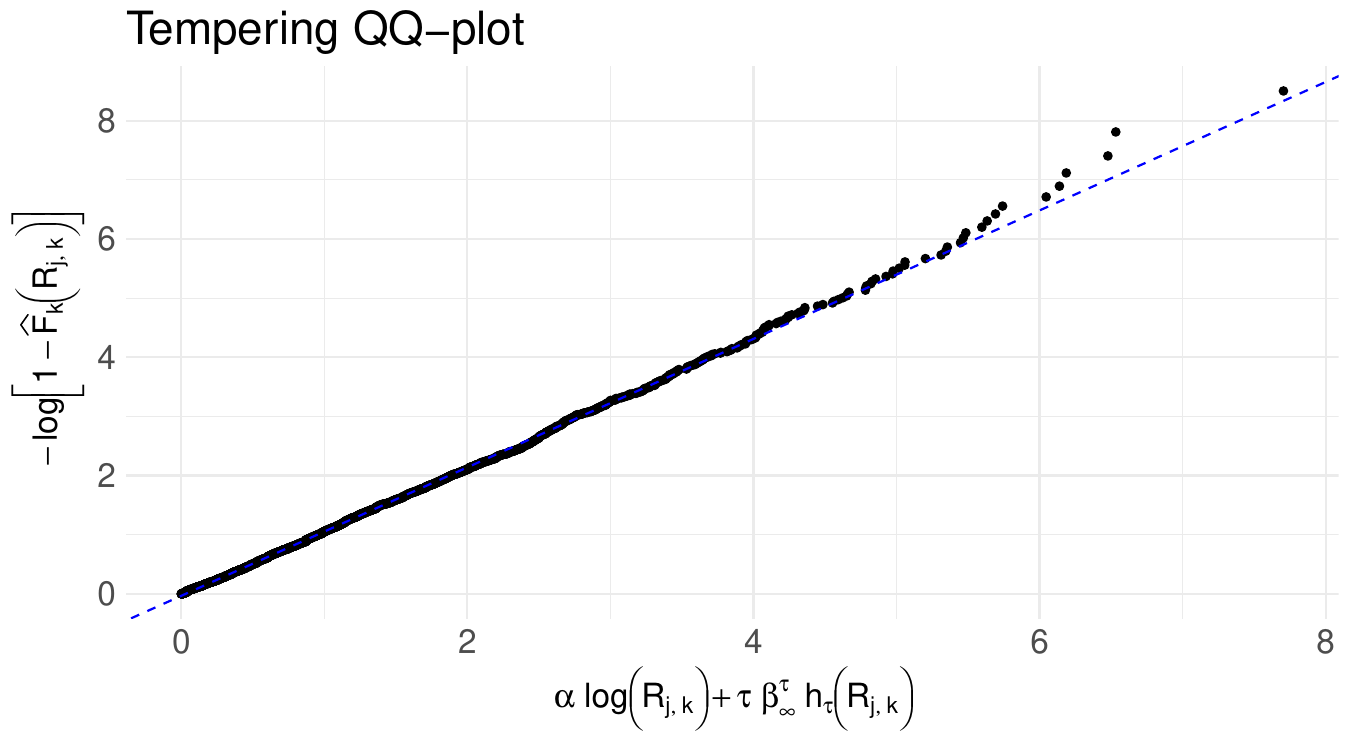}}
	\caption{QQ-plots for Norwegian fire claims for Pareto, Weibull and tempering fit using the points in \eqref{eqline} .}\label{norwegian}
\end{figure}

\noindent{\bf Example: Norwegian Fire Insurance Data.} Here, we analyze the Norwegian fire insurance data as discussed in Albrecher et al.\ \cite{AAB_Astin21}. Figure \ref{norwegian} depicts the  Pareto  and Weibull QQ-plots (the latter for the 100 largest data points). The Pareto QQ-plot becomes concave near the top points where a Weibull model appears to fit. Hence this data set is a candidate for the proposed tempering model. In Figure \ref{norwegian} (bottom), we present the plot of the points in \eqref{eqline} with $\hat{\alpha}=1.19928,\, \hat{\beta}_{\infty}= 0.003958,\, \hat{\tau}= 0.70227$ obtained at the optimal value  $\hat{k}=4920$, which is linear overall.  

\section{Censoring}\label{sec3}
In various lines of insurance business (so-called \textit{long-tailed business}) one faces long periods until claim sizes are finally settled, e.g., in  liability lines. At the time of the evaluation of the portfolio, a proportion of the claims are not fully dealt with yet, and hence the real cumulative
payments are unknown and censored by the payments up to evaluation. Censoring can also be due to policy limits which conceal the real underlying loss amounts. When using extreme value techniques in the risk assessment, it is important to take these censoring mechanisms into account in an appropriate way.\\

To formalize such a setting, let us assume that the loss random variable $X$ of interest is Pareto-type distributed with extreme value index $\xi>0$, and let $C$ be a censoring random variable that is also Pareto-type distributed, possibly with a different extreme value index.  One considers  $Z = \min(X, C)$ and an indicator $\delta = {\mathbb{I}}(X \leq C)$ which equals 1 if the observation $Z$ is non-censored.\\
Let the ordered $Z$-observations and the corresponding $\delta$-values be denoted by $\{(Z_{(i)},\delta_{(i)})\}_{i=1}^n$. Then, based on the POT principle, the classical censoring likelihood based on the relative excesses $R_{j,k} = Z_{(n-j+1)}/Z_{(n-k)}$, $j = 1, \ldots, k$ is given by
\[
\prod_{j=1}^k (\xi^{-1} R_{j,k}^{-\xi^{-1}-1})^{\delta_{(n-j)} }(  R_{j,k}^{-\xi^{-1}})^{1-\delta_{(n-j+1)}}.
\]
Optimizing with respect to $\xi$ leads to the following Hill-type estimator adapted for censoring:
\begin{equation}\label{hillc}
H_{k,n}^{(c)}= \frac{k^{-1}\sum_{j=1}^k \log \left(Z_{(n-j+1)}/Z_{(n-k)}\right) }{k^{-1}\sum_{j=1}^k 
\delta_{(n-j+1)}}.
\end{equation}
This estimator equals the original Hill estimator based on the $Z$-observations, divided by the proportion of non-censored data among the top $k$ observations.\\
In case of random right-censoring, and assuming independence of the $X$ and $C$ sequences, the Kaplan–Meier product-limit estimator $\hat{F}_n$ is the nonparametric maximum likelihood estimator of the distribution function of $X$, given for all $x\in(0,\infty)$ by
$$
1-\hat{F}_n (x) = \prod_{i=1}^n \left(1-\frac{\mathbb{I}(Z_{(i)}\leq x)}{n-i+1} \right)^{\delta_{(i)}}.
$$
Beirlant et al.\ \cite{BGDF07} showed that $H_{k,n}^{(c)}$ can also be obtained as a slope estimator of the
Pareto QQ-plot adapted for censoring, featuring the points
\[
\left\{(  -\log \{1-\hat{F}(Z_{(n-j+1)})\}, \log Z_{(n-j+1)} )\right\}_{j=1}^n.
\]
 Worms and Worms\ \cite{WW2016} proposed
$$
\hat{\xi}_k^W = \sum_{j=1}^k \frac{\hat{F}_n (Z_{(n-j+1)})}{\hat{F}_n (Z_{(n-k)})} \left(\log Z_{(n-j+1)}- \log Z_{(n-j)}\right)
$$
as an estimator of $\int_1^\infty {\overline{F}(ut)}/{\overline{F}(t)} \diff (\log u) \to \xi$ as $t \to \infty$, with $F$ denoting the distribution function of $X$.
% This estimator has been shown to improve significantly over $H_{k,n}^{(c)}$. Einmahl et al.\ 
\cite{einmahl2008statistics} obtained a moment estimator  $\hat{\xi}_k^M$ under censoring by simple division of the moment estimator by $k^{-1}\sum_{j=1}^k 
\delta_{(n-j+1)}$. This estimator is then consistent, not only for Pareto-type distributions, but for $F$ belonging to any max-domain of attraction. They also provided the first asymptotic results for the available estimator, while Beirlant et al.\ \cite{beirlant2019estimation} provided asymptotic results for a version close to $\hat{\xi}_k^W$.  Other estimators were proposed in  Bladt et al.\ \cite{bladt2021trimmed}. Recently, Bladt and Rodionov \cite{bladt2023censored} proposed the analysis of integrals based on the product-limit
estimator of normalized top-order statistics. These integrals allow to derive asymptotic distributional properties, offering an alternative approach to
conventional plug-in estimation methods.\\

%  The asymptotic results of the moment estimator proposed by Einmahl et al.\ \cite{einmahl2008statistics} are revisited. \\
The available estimators tend to have a larger bias that increases with the proportion of intermediate data that are censored. Bias-reduced methods are proposed in  Beirlant et al.\ \cite{BBdWG2016,beirlant2018penalized} and Bladt et al.\ \cite{bladt2021trimmed}, among others. \\

When censoring occurs due to long development times so that the final claim amount of a larger proportion of claims is underestimated at the evaluation time of the portfolio, caution is advised when using the available asymptotic normality results since the independence of $X$ and $C$ can not be taken for granted. Indeed, larger claims have a larger probability of being censored due to a longer development time. Stupfler \cite{stupfler2019} is a first paper to address this issue. \\ 

Extreme quantile estimators have been proposed in several of the abovementioned papers adapting the Weissman\ \cite{weissman1978} estimator to the censoring case:
$$
\hat{Q}_k^{(c)}(1-p) = Z_{(n-k)}\left\{{1-\hat{F}_n(Z_{(n-k)}) \over p} \right\}^{\hat{\xi}_k^{(c)}},
$$ 
where $\hat{\xi}_k^{(c)}$ denotes any of the extreme value index estimators discussed above.\\

\begin{figure}[htb]
\centerline{\includegraphics[height=160pt]{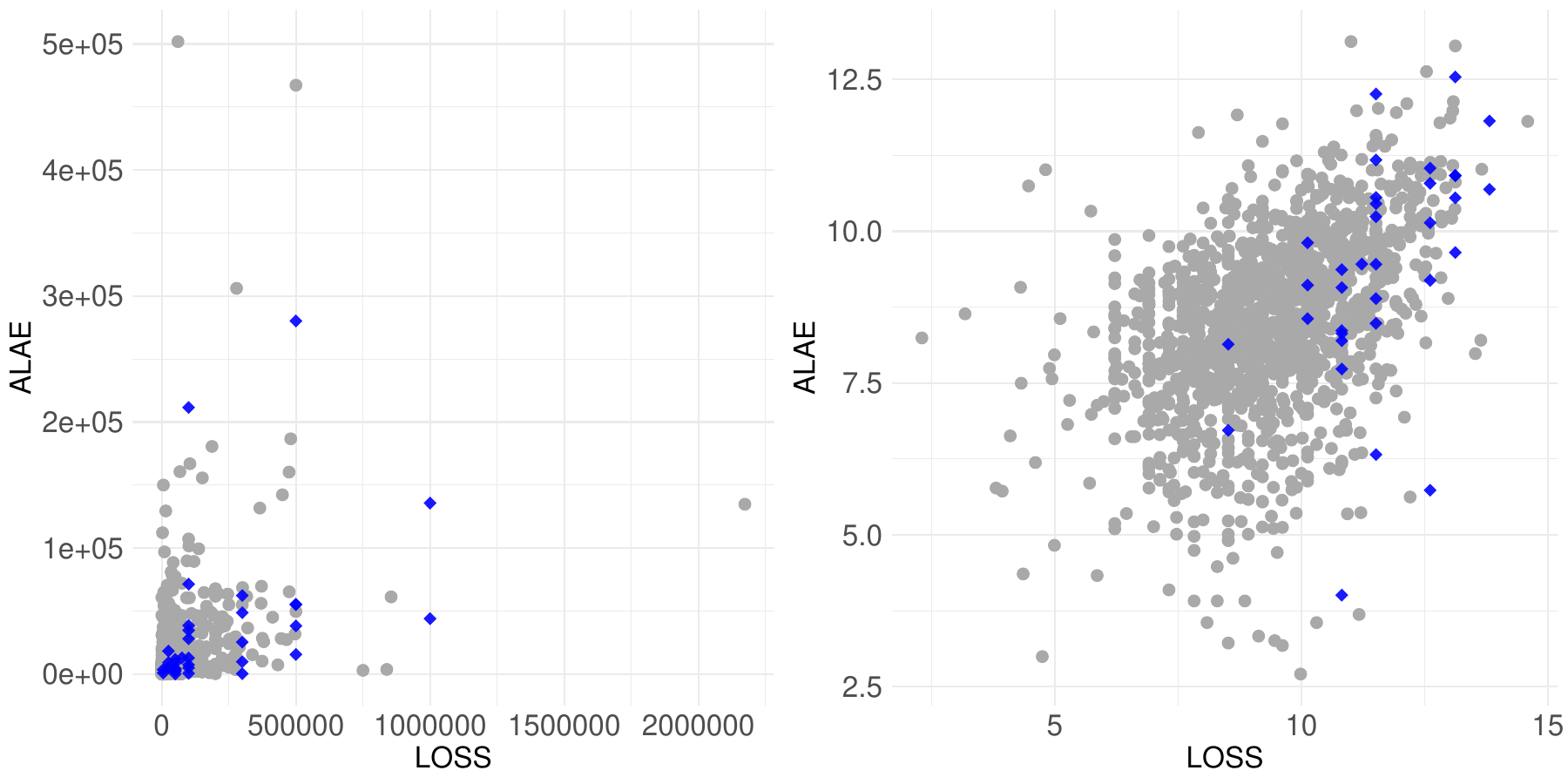}}
	\caption{Loss versus ALAE data. Original scale (left) and $\log$-scale (right).}\label{alae1a}
\end{figure}

\begin{figure}[htb]
	\centerline{\includegraphics[height=160pt]{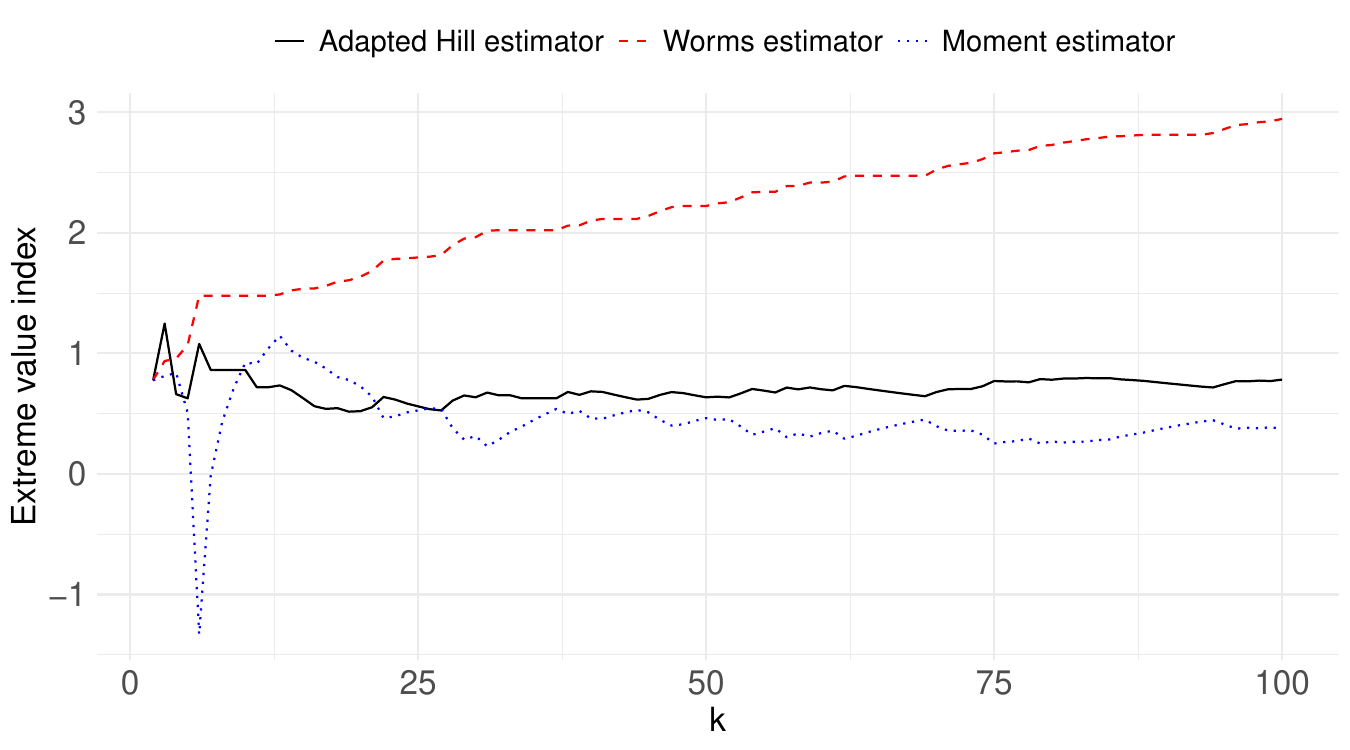}}
	\caption{Loss data: Comparison of extreme value index estimators $H_{k,n}^{(c)}$, $\hat{\xi}_k^W$, and $\hat{\xi}_k^{M}$ as a function of $k$.}\label{alae2a}
\end{figure}

\noindent
{\bf Example: Censored Liability Loss Data.} Goegebeur et al.\ \cite{GGQ_IME19} discussed the (Loss, ALAE) data set from the R package \texttt{copula} containing 1500 general liability claims with indemnity claims (Loss) and allocated loss adjustment expense (ALAE) related to the settlement of individual claims, such as expenses for lawyers or claim investigations. One finds that 34 observations have censored losses. In Figure \ref{alae1a}, the data are plotted with the censored data in blue. Figure \ref{alae2a} displays $H_{k,n}^{(c)}$ (solid bold line), $\hat{\xi}_k^W$ (dashed line) and $\hat{\xi}_k^M$ (dotted line) as a function of $k$. These estimates indicate a heavy tail with an extreme value index ranging between 0.5 and 1. $\hat{\xi}_k^W$ appears to suffer substantial  bias which only decreases with smaller $k$. 

In cases  with a high percentage of right-censored data, one can also rely on available expert information for the censored observations. This setting arises for instance for liability insurance claims, where actuarial experts build reserves based on the specificity of each open claim. These reserves can be used to improve estimation based on the already available data points from closed claims. The incorporation of expert opinion is discussed in  Bladt et al.\ \cite{BAB_SAJ20}.  Albrecher and Bladt \cite{albrecher2022informed} extend the statistical censoring setup to the situation when
random measures can be assigned to the realization of datapoints, leading to a
new way of incorporating expert information into the usual parametric estimation
procedures.  In some types of insurance, such as  motor third-party liability insurance and both
worker’s compensation and loss of income coverage, claims might potentially reopen.
 Bladt and Furrer \cite{bladt2023expert} propose adaptations to the Kaplan\textendash Meier function in order to incorporate such contaminations.

\section{Full Models for Claims}\label{sec5}
Actuaries use models for claim sizes to set premiums,
calculate risk measures and determine capital requirements for solvency regulations. Extreme value models, based for instance on the generalized Pareto distribution, are not able to capture the characteristics over the entire range of the loss distribution, which makes them unsuitable as a global model. It is often imperative to obtain
an overall fit for the distribution of losses, for example in a risk analysis where the focus is
not only on extreme events, or when setting up a reinsurance program. Instead of trying
many different standard distributions, \textit{splicing} two distributions (Klugman et al.\ \cite{klugman2012loss}) is
often more suitable to model the complete loss distribution. In the literature, a splicing model is
also called a \textit{composite model}.\index{composite model} We hereby combine an analytically tractable (typically light-tailed) distribution for the body
which covers small and moderate losses (the so-called \textit{losses!attritional}), with a heavy-tailed
distribution for the tail to capture the \textit{losses!large}. In the actuarial literature, simple splicing
models have been proposed. Beirlant et al.\ \cite{beirlant2004statistics} and Klugman et al.\ \cite{klugman2012loss} consider the
splicing of the exponential distribution with the Pareto distribution. Other distributions
for the body such as the Weibull distribution (Ciumara \cite{ciumara2006actuarial}; Scollnik and Sun \cite{scollnik2012modeling}) or
the lognormal distribution (Cooray and Ananda \cite{cooray2005modeling}; Scollnik \cite{scollnik2007composite}; Pigeon and Denuit \cite{pigeon2011composite}) have also been used. Nadarajah and Bakar \cite{nadarajah2014new}, Bakar et al.\ \cite{bakar2015modeling}, and  Calder\'in-Ojeda and Kwok \cite{calderin2016modeling} investigate the splicing of the log-normal or Weibull distribution
with various tail distributions. Lee et al.\ \cite{lee2012modeling} consider the splicing of a mixture of two exponentials and the GP distribution. The use of a mixture model in the first splicing component
gives more flexibility in modelling the small and moderate losses, see Fackler \cite{fackler2013reinventing}. For a splicing approach to model cyber risk data, see, e.g., \cite{eling2019actual}.\\

Rather than a complex search for appropriate splicing combinations of extreme and attritional models,  Reynkens et al.\ \cite{reynkens2017modelling} proposed a semi-automatic method for splicing, a mixed Erlang (ME) \index{mixed Erlang distribution}
with density 
\[
f_{\text{ME}}(x; {\bf r}, {\bf \alpha},\lambda) = \sum_{j=1}^M \alpha_j {\lambda^{r_j} \over (r_j-1)!} x^{r_j-1}e^{-\lambda x},\, x>0
\]
and integer parameters ${\bf r} =(r_1,\ldots,r_M)$ with $r_1 < \cdots < r_M$, and 
%
%and survival function
%\[
%1-F_{ME}(x; {\bf r}, {\bf \alpha},\lambda) = e^{-\lambda x} \sum_{j=1}^M \alpha_j \sum_{n=0}^{r_j -1}{(\lambda x)^{n} \over n!} ,
%\]
an extreme value model such as a Pareto, generalized Pareto, or even truncated or tempered versions of those. Here, $(\alpha_1, \ldots, \alpha_M)$ with $\alpha_j>0$ and $\sum_{j=1}^M \alpha_j =1$ are the weights in the Erlang mixture. 
%The Erlang density is given by
%\[
%f_E(x; r,\lambda) = {\lambda^r \over (r-1)!} x^{r-1}e^{-\lambda x}, \; x>0,
%\]
%where $r$ is a positive integer shape parameter. Following  we consider mixtures of $M$-Erlang distributions with common scale parameter $1/\lambda$ having density
%
%where the positive integers ${\bf r} = (r_1,\ldots,r_M)$ with $r_1 < r_2<\ldots < r_M$ are the shape parameters of the Erlang distributions, and 
The class of mixtures of Erlang distributions with a common scale $1/\lambda$ is dense in the space of distributions on $\mathbb{R}^+$, and it is closed under mixtures, convolution and compounding. Hence aggregate risk calculations are simple, and XL premiums and risk measures based on quantiles can also be evaluated in a rather straightforward way. For instance, a composite ME-generalized Pareto distribution for some $0<\pi<1$ has density
\begin{equation*}
f_{\text{ME,GP}}(x) =
\left\{
\begin{array}{ll}
\pi\frac{f_{\text{ME}}(x; {\bf r}, {\bf \alpha},\lambda)}{F_{\text{ME}}(t; {\bf r}, {\bf \alpha},\lambda)}, & 0 <x \leq t, \\
(1-\pi) {1 \over \sigma}\left\{ 1+{\xi \over \sigma}(x-t) \right\}^{-1-1/\xi},    & x>t,
\end{array}
\right.
\end{equation*}
and survival function 
\begin{equation*}
1-F_{\text{ME,GP}}(x) =
\left\{
\begin{array}{ll}
1-\pi\frac{F_{\text{ME}}(x; {\bf r}, {\bf \alpha},\lambda)}{F_{\text{ME}}(t; {\bf r}, {\bf \alpha},\lambda)}, &  0<x \leq t, \\
(1-\pi) \left\{ 1+{\xi \over \sigma}(x-t) \right\}^{-1/\xi},    & x>t.
\end{array}
\right.
\end{equation*}
%One can for instance choose $\hat\pi = 1-{k_*/n}$, where $k_*$ is an appropriate number of top order statistics corresponding to an extreme value threshold $t = x_{n-k_*,n}$. 
%\\

Fitting ME distributions through direct likelihood maximization is difficult. Lee and Lin \cite{lee2010modeling} use the {\it Expectation-Maximization} (EM) algorithm proposed by Dempster et al.\   \cite{dempster1977maximum} to fit the ME distribution. Model selection criteria, such as the AIC and BIC information criteria, are then used to avoid overfitting. Verbelen et al.\ \cite{verbelen2015fitting} extend this approach to censored and/or truncated data. 
%
%The need for the EM algorithm follows from the data incompleteness due to mixing and censoring. It is used to compute the MLE for incomplete data where direct maximization is impossible. It consists of two steps  that are put in an iteration until convergence:
%\begin{itemize}
%\item {\bf E-step}: compute the conditional expectation of the log-likelihood given the observed data and previous parameter estimates;
%\item {\bf M-step}: determine a subsequent set of parameter estimates in the parameter range through maximization of the conditional expectation computed in the E-step.
%\end{itemize} 

Rather than proposing a data-driven estimator of the splicing point $t$, one can use an expert opinion on the splicing point $t$ based on EVA as outlined above. Then, $\pi$ can be estimated by the fraction of the data not larger than $t$.  The extreme value index $\xi$ is estimated in the algorithm, starting from the value obtained from the EVA at the threshold $t$. The final estimates for $\xi$ turn out to be close to the EVA estimates. Next, the ME parameters $(\alpha, \lambda)$ are estimated using the EM algorithm, cf.\ \cite{verbelen2015fitting}. The number of ME components $M$ is estimated using a backward stepwise search, starting from a certain upper value, whereby the smallest shape is deleted if this decreases an information criterion\index{information criterion} such as AIC or BIC. Moreover, for each value of $M$, the shapes ${\bf r}$ are adjusted based on maximising the likelihood starting from ${\bf r} = (s, 2s, \ldots, M \, s)$, where $s$ is a chosen spread.
\\

\begin{figure}[htb]
	\centerline{
	\includegraphics[height=160pt]{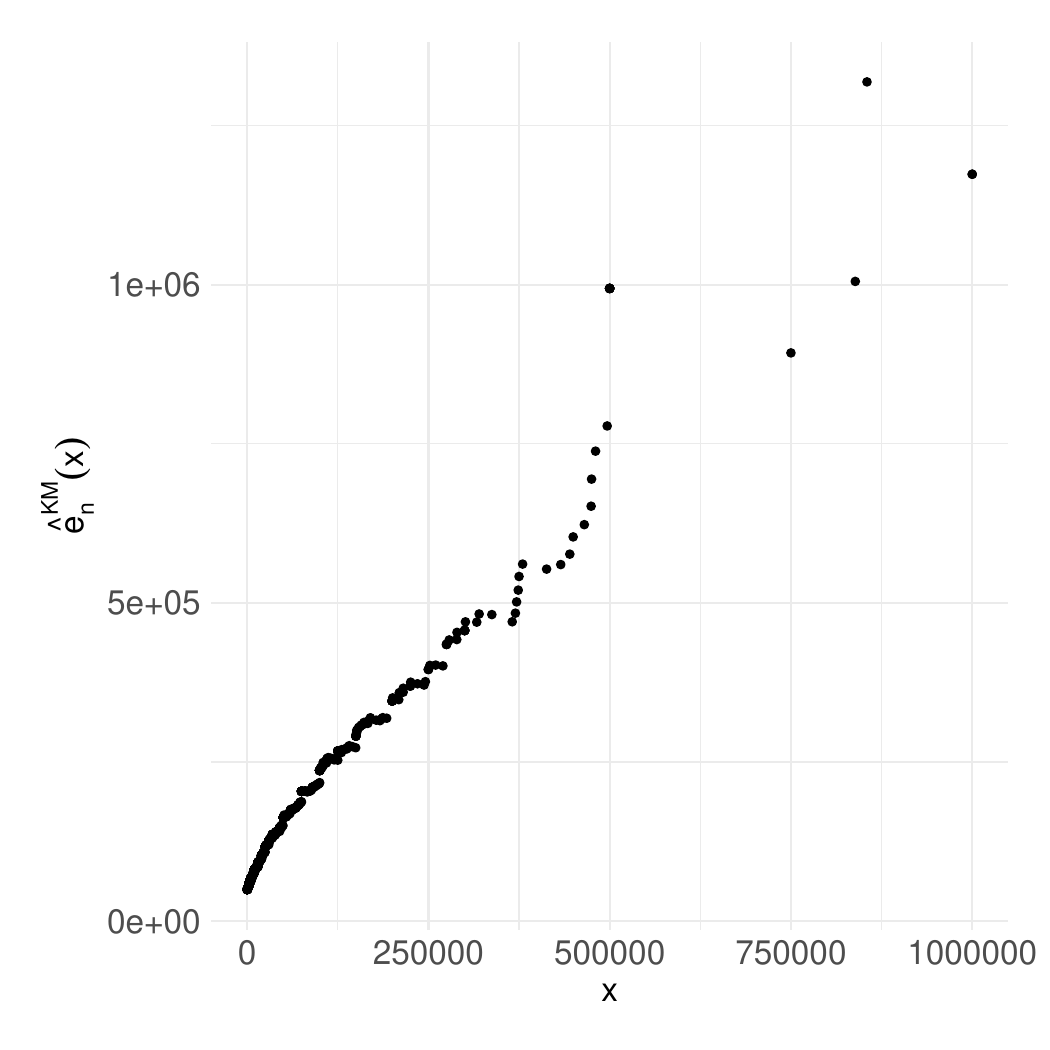}
	\hspace{1cm}
	\includegraphics[height=160pt]{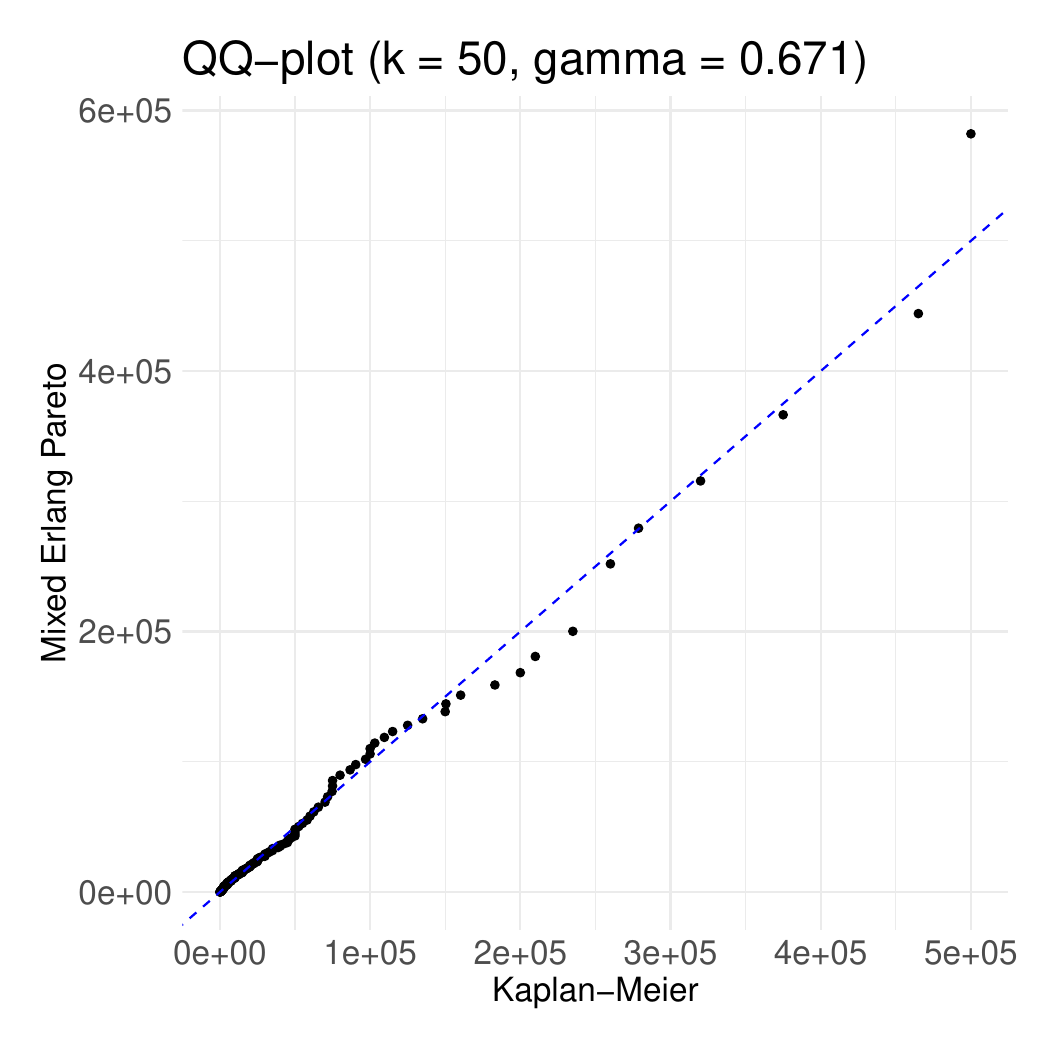}
	}
	\caption{Loss variable from (Loss, ALAE) data set. Mean excess function $\hat{e}_n^{\text{KM}}$ (left) and splicing QQ-plot (right).}\label{alaesplice1}
\end{figure}

\noindent
{\bf Example: Censored Liability Loss Data.} We continue the analysis of the Loss variable from the (Loss, ALAE) data set considered above. To search for splicing points, estimates of the mean excess function\index{mean excess function}
$$
e(x):= { E}\left( X-x \mid X>x\right)= 
{\int_x^{\infty} \overline{F}(u) \diff{u} \over \overline{F}(x)}
$$
are found to be useful. Substituting $\overline{F}$ with the Kaplan\textendash Meier survival function leads to an estimator $\hat{e}_n^{\text{KM}}(x)$ as a function of $x$, displayed in Figure \ref{alaesplice1} (left). A concave increase appears to change into a linear pattern, starting at a threshold level around $10^5$.  \\ 

 In Figure \ref{alaesplice1} (right), the QQ-plot is given with the Kaplan\textendash Meier quantiles set out against the quantiles of the fitted ME-Pareto splicing model.  The splicing point is taken  at the 50th largest observation, leading to a good fit with $\hat{\xi}= 0.67$ and three ME components. \\

When estimating $\mbox{VaR}_{1-p}$ for a two-component spliced distribution, we have 
\begin{equation*}
Q(1-p) = \begin{cases}
Q_1((1-p)/\pi), & \quad \text{if } 0 \leq p \leq \pi, \\
Q_2((1-p-\pi)/(1-\pi))= Q_2 \left(1- {p \over 1-\pi}\right),  & \quad \text{if } \pi < p \leq 1,
\end{cases} 
\label{VaRSplice}
\end{equation*}
where $Q_1$ denotes the quantile function of the ME component and $Q_2$ of the tail component. $Q_1$ can be obtained numerically. 
%In case the tail component is given by a simple Pareto distribution, we have 
%\begin{equation*}
%Q_2(1-u) = t \; u^{-\xi},\; 0<u<1,
%\end{equation*}
Taking $t=X_{(n-k)}$ and $1-\pi = (k+1)/(n+1)$, 
we are led to the appropriate extreme quantile estimator as discussed above,   when $\pi < p \leq 1$.
%Using  an upper-truncated Pareto, respectively a generalized Pareto, tail fit one of course can use $ \hat{q}^{T}_{k,p}$, respectively $ \hat{q}^{ML}_{k,p}$  in case $\pi < p \leq 1$. 
\\

When estimating $\Pi (u)$ as defined in \eqref{tthat}, we again identify two cases:  $u \leq t = X_{(n-k)}$, or $u > t= X_{(n-k)}$  in which case statistics of extremes can be used. When $u > t$, then 
\begin{eqnarray*}
\Pi(u) &=& \int_u^{\infty} [1-\{\pi+(1-\pi)F_2(z)\}]\diff{z} \\& =& (1-\pi) \int_u^{\infty} \{1-F_2(z)\} \diff{z} =:(1-\pi)\Pi_2(u).
\end{eqnarray*}
When $u < t$, we have 
\begin{align*}
\Pi(u) &= \int_u^t \{1-\pi F_1(z)\} \diff{z} + \int_t^{+\infty} [1-\{\pi+(1-\pi)F_2(z)\}] \diff{z} \\
&= (t-u) - \pi \int_u^t  F_1(z) \diff{z} + (1-\pi) \int_t^{+\infty} \{1-F_2(z)\}\diff{z} \\
& =(t-u) - (t-u)\pi + \pi \int_u^t \{1-F_1(z)\} \diff{z} + (1-\pi)\Pi_2(t)\\
& =(1-\pi) (t-u) + \pi \Pi_1(u) + (1-\pi)\Pi_2(t).
\end{align*}
%Note that $\Pi(u)=0$ for $u\geq T$ and $\Pi(u)=\Pi(t^l)+(t^l-u)$ for $u\leq t^l$.
Let $t^l$ denote the lower limit of the support of the distribution of $X$, and $T$ the upper limit.
For the ME distribution, we get
\begin{eqnarray*}
\Pi_1(u) &=& \int_u^t \left\{1- \frac{F_1^*(z)-F_1^*(t^l)}{F_1^*(t)-F_1^*(t^l)} \right\} \diff{z} \\
 &=& \frac{\left\{F_1^*(t)-1\right\}(t-u) + \{\Pi_1^*(u)-\Pi_1^*(t)\}}{F_1^*(t)-F_1^*(t^l)},
\end{eqnarray*}
with
\begin{equation*}
F_1^* (x) = \sum_{j=1}^M \alpha_j \left( 1-\sum_{n=0}^{r_j -1}
e^{-\lambda x}{(\lambda x)^{n} \over n!}
\right)
\end{equation*}
and, assuming that $r_n=n$ for $n=1,\ldots,M$,
\[
\Pi^{*}_1 (u) = {1 \over \lambda} e^{-\lambda u}\sum_{n=0}^{M-1}\sum_{k=n}^{M-1}
\left(\sum_{j=k+1}^{M} \alpha_j \right)
\frac{(\lambda u)^n}{n!}.
\]
For $\Pi_2(u)$ in case $u>t=X_{(n-k)}$, see for instance Goegebeur et al. \cite{goegebeur2023conditional} for censored data. 
More recent papers on splicing methods are Li and Liu \cite{li2023claims}, Bladt and Yslas \cite{bladt2022heavy},  Poudyal and Brazauskas \cite{poudyal2023finite}, Wang et al. \cite{wang2020modelling}, Raschke\ \cite{raschke2020alternative}, Wang and Hob{\ae}k-Haff \cite{wang2019focussed}, Bolviken and Hob{\ae}k-Haff \cite{bolviken2024loss}, Fung et al.\ \cite{fung2023soft} and Ghaddab et al.\ \cite{ghaddab2023extreme}. \\

While the splicing approach allows to disentangle the fitting of the tail and the body of the claim size distribution, it may not seem natural to glue together two (or even more) different distributions. Also, the choice of the splicing point $t$, where one transits from one `regime' to another remains somewhat arbitrary. At the same time, flexible and mathematically tractable families of distributions like phase-type distributions (of which the ME family above is a special case) are not well-suited for heavy tails because their tail is exponentially bounded and many components in the mixture (i.e.\ a high phase-type dimension) would be needed to provide reasonable fits for higher quantiles, and even then the tail behavior is not naturally captured (which is one reason why the ME above is not used for the entire range of the distribution). Another recent alternative that circumvents the threshold choice of splicing is the class of matrix-Mittag\textendash Leffler distributions and its power transforms, cf.\ Albrecher et al.\ \cite{MML20}. This class results from inhomogeneous phase-type distributions and can be seen as transforms of matrix-exponential distributions leading to heavy tails, essentially comprising Mittag\textendash Leffler distributions with matrix-valued parameters. In this way, one avoids the need for an increased number of parameters to compensate for too light tails with yet tractable expressions for both the fitting and subsequent calculations. The quality of the resulting fit across the entire range of the distribution can be quite remarkable, see Albrecher et al.\ \cite{MML20} and Bladt and Yslas \cite{bladt2022heavy} for details.

%Many specific challenges, such as (see later section titles)

\section{Regression Modelling}\label{sec6} \index{regression modelling}
In recent years, advanced data collection and storage techniques and practices have significantly increased the potential for relevant covariate information for the occurrence of extremes. This is of course also very relevant for insurance applications. We focus here on adaptations of techniques for censored data when covariate information is available. Regression modelling of a censored random variable $Y$ as a function of a covariate $x$ is available at specific covariate values using local smoothing methods. For instance, Albrecher et al. \cite[Sec. 4.4.3]{ABT17} propose extensions of the above univariate estimation methods under random right censoring,   based on the Akritas\textendash Van Keilegom \cite{akritas2003estimation} extension of the Kaplan\textendash Meier estimator.\index{Kaplan\textendash Meier estimator} They use this method to model the extremes of final claim payments in a long-tailed portfolio as a function of the number of development years. In this way, a positive dependence between development times and heaviness of the tail can be  detected.\\

Considering a univariate covariate $X$ with a specific value $x$, and  assuming that $Y$ and the censoring variable $C$ are conditionally independent given $X$, we have
\[1-\hat{F}_{Y|X}(y\mid x) = \prod_{Z_{i} \leq y } \left(1-\frac{W_{i}(x;h)}{\sum_{j=1}^n W_{j}(x;h) \mathbb{I}(Z_{j} \geq Z_{i})}\right)^{\delta_i}\]
with weights
\[W_{i}(x;h) = \begin{cases}
	K\left(\frac{x-x_{i}}{h}\right)/\sum_{\delta_j=1}K\left(\frac{x-x_{j}}{h}\right), & \qquad \text{if } \delta_i=1,\\
	0, & \qquad \text{if } \delta_i=0,
\end{cases}\]
where $Z=\min (Y, C)$, $K$ is a kernel function and $h$ is the chosen bandwidth.
Denoting the weight $W$ corresponding to the $i$th smallest $Z$-value
$Z_{i,n}$ by $W_{i,n}$, we for instance obtain the following Worms-type estimator of the conditional extreme value index given $X=x$: 
\begin{eqnarray*}
	\hat{\xi}_{k,n}^{W}(X=x)
	&=&   \frac{\int_{Z_{n-k,n}}^{\infty} \{1-\hat{F}_{Y|X}(y\mid x) \}\, \diff (\log y)} {1-\hat{F}_{Y|X}(Z_{(n-k)}\mid x)} \\
	&=& 
	\frac{\sum_{j=1}^k \left(\prod_{i=1}^{n-j}\left[ \left(1-\frac{W_{i,n}(x;h)}{1-\sum_{l=1}^{i-1} W_{l,n}(x;h)} \right)^{\delta_{i,n}}\right] \log {Z_{(n-j+1)}\over Z_{(n-j)}}\right)}{\prod_{i=1}^{n-k}\left[ \left(1-\frac{W_{i,n}(x;h)}{1-\sum_{l=1}^{i-1} W_{l,n}(x;h)} \right)^{\delta_{i,n}}\right]}. 
\end{eqnarray*}
Pareto QQ-plots adapted for censoring per  chosen $x$ value, defined as
\[
\left\{(  -\log \{1-\hat{F}_{Y|X}(Z_{(n-j+1)}\mid x)\}, \log Z_{(n-j+1)} )\right\}_{j=1}^n,
\]
can then help to assess an appropriate value of $k=k_x$. 
\\
A generalization of extreme quantile estimators $\hat{Q}_k^{(c)}(1-p)$ to the local regression setting is obtained using a local extreme value index estimator such as $\hat{\xi}_{k,n}^{W}(X=x)$ and the Akritas\textendash Van Keilegom estimator $1-\hat{F}_{Y|X}(y\mid x)$. \\

Stupfler\ \cite{stupfler2016} presents a local  estimator of the extreme value index under random right censoring when $Y$ given $X=x$ belongs to any max-domain of attraction, generalizing the moment estimator proposed in Einmahl et al.\ \cite{einmahl2008statistics}. Further extensions were provided in Goegebeur et al.\
\cite{GGQ_IME20}, Rutikanga and Diop \cite{rutikanga2021functional} and Dierckx et al.\ \cite{dierckx2021local}.

%\noindent
%??Example:  windspeed Austria example??  \\

%\noindent
%Regression extremes with censored data. \cite{GGQ_IME20} 

\section{Multivariate Modelling}\label{sec62} 
The consideration of multivariate extremes is relevant for reinsurance companies in many ways. Firstly, it naturally appears when it comes to aggregate covers across different business lines of a first line insurer, but then also for the purpose of spatial diversification across contracts with several insurers, and more generally in the overall resulting reinsurance portfolio. The general mathematical theory of multivariate extremes is a well-developed and vibrant field (see e.g.\ Beirlant et al.\ \cite{beirlant2004statistics}, Davison and Huser \cite{davison2015statistics} and Engelke and Ivanovs \cite{engelke2021sparse} for overviews). We restrict ourselves here to some contributions that are particular to insurance applications, i.e.\ extensions of the previous sections to the multivariate situation.\\ 

Lee and Lin \cite{lee2012modeling} define a $D$-variate Erlang mixture, where each mixture component is the joint distribution of $D$ independent Erlang distributions with a common scale parameter $1/\lambda >0$. The dependence structure is then  captured by the combination of the positive integer shape parameters of the Erlangs in each dimension. For illustration, we entirely focus on the case $D=2$. Let ${\bf r}=(r_1,r_2)\in \mathcal{R}$ the vector of shape parameters with values in the set $\mathcal{R}$ of allowed combinations in the set of all shape vectors with non-zero weight. The density of a bivariate Erlang mixture\index{mixed Erlang distribution} evaluated in ${\bf x}=(x_1,x_2)>{\bf 0}$ is then defined as
\begin{equation*}
	f_{\text{MME}}({\bf x}; {\bf \alpha},{\bf r}, \lambda)=
%	\sum_{{\bf r} \in \mathcal{R}}\alpha_{{\bf r}}\, \prod_{j=1}^{2} f_E(x_j, r_j,\lambda)=
	\sum_{{\bf r} \in \mathcal{R}}\alpha_{{\bf r}}\, \prod_{j=1}^{2} 
	{\lambda^{r_j}x_j^{r_j-1}e^{-\lambda x_j} \over (r_j-1)!}.
	\label{MME}
\end{equation*}
It can be shown that the bivariate Erlang mixture with density
\[
%f_{MME}({\bf x};  \lambda)= 
\sum_{r_1=1}^{\infty} \sum_{r_2=1}^{\infty} \alpha_{{\bf r }} (\lambda)\, \prod_{j=1}^2 {\lambda^{r_j}x_j^{r_j-1}e^{-\lambda x_j} \over (r_j-1)!}
\]
and suitable mixing weights can approximate any given bivariate positive distribution function arbitrarily closely as $\lambda \to \infty$, 
%\[
%\alpha_{{\bf r }} = 
%\int_{(r_1-1)/\lambda}^{r_1/\lambda}  \int_{(r_2-1)/\lambda}^{r_2/\lambda} f({\bf x})d{\bf x}
%\]
%satisfies $\lim_{\lambda \to \infty} F_{MME}({\bf x};  \lambda)= F({\bf x})$, 
cf.\ \cite{lee2012modeling}. 
%The weights $\alpha_{{\bf r }}$ of the components in the mixture are defined by integrating the density over the corresponding $d$-dimensional rectangle of the grid formed by the shape parameters multiplied with the common scale. When the value of $\lambda$ increases, this grid becomes more refined and the sequence of Erlang mixtures converges to the underlying  distribution function. 
%\\
Verbelen et al.\ \cite{verbelen2015fitting} provide a flexible fitting procedure for such multivariate mixed Erlang (MME) distributions, which iteratively uses the EM algorithm, by introducing a computationally efficient initialization and adjustment strategy for the shape parameter vectors. \cite{verbelen2015fitting} also considered  randomly censored and fixed truncated data.\\

When it comes to splicing, the bivariate cumulative distribution function of a  splicing model with a bivariate ME and a bivariate GP distribution is now 
\begin{equation} \label{MEGPD2}
	F({\bf x}) =
	\left\{
	\begin{array}{ll}
		0, & \mbox{if } {\bf x} \leq {\bf t}^l, \\
		\pi F_{\text{MME}}({\bf x}), & \mbox{if } {\bf t}^l \leq {\bf x} \leq {\bf t}, \\
		\pi F_{\text{MME}}({\bf x}) + (1-\pi)F_{\text{MGPD}}({\bf x}), & \mbox{if }  {\bf x} \nleq {\bf t},
	\end{array}
	\right.
\end{equation}
for an appropriate set of thresholds ${\bf t}=(t_1,t_2)$, see also \cite{ABT17}. 
%, with corresponding distribution function 
%\[
%F_1({\bf x}; {\bf t}^l,{\bf t}) = \begin{cases}
%	0 & \text{if } {\bf x}\ngeq {\bf t}^l \\
%	\frac{\sum_{{\bf r}\in \mathcal{R}} \alpha_{{\bf r}} \prod_{j=1}^2 F_E(\min\{x_j,t_j\}; r_j, \lambda)}{\sum_{{\bf r}\in \mathcal{R}} \alpha_{{\bf r}} \prod_{j=1}^2 (F_E(t_j; r_j, \lambda)-F_E(t^l_j; r_j, \lambda))}  & \text{if } {\bf x} \geq {\bf t}^l \text{ and }{\bf x} \ngeq {\bf t}\\
%	1 & \text{if } {\bf x} \geq {\bf t}.
%\end{cases}
%\]
For the tail component one can for instance use the bivariate GP distribution as proposed in Rootz\'en and Tajvidi \cite{rootzen2006multivariate} and Kiriliouk et al.\ \cite{kiriliouk2019peaks}:
\begin{eqnarray*}
	F_{\text{MGPD}}({\bf x}) &= \theta^{-1}\left[
	\ell\left( \left\{1+{\xi_1 ((x_1-t_1)\wedge 0)\over \sigma_1}\right\}^{-1/\xi_1},\left\{1+{\xi_2 ((x_2-t_2) \wedge 0) \over \sigma_2} \right\}^{-1/\xi_2}\right) \right. \\
	& \hspace{1cm}\left.
	-\ell\left(\left\{1+{\xi_1 (x_1-t_1)\over \sigma_1}\right\}^{-1/\xi_1},\left\{1+{\xi_2 (x_2-t_2) \over \sigma_2} \right\}^{-1/\xi_2}\right)\right],
\end{eqnarray*}
such that ${\boldsymbol \sigma} + {\boldsymbol \xi}({\bf x-t}) > {\bf 0}$, where $\ell$ denotes the stable tail dependence function and $\theta=\ell(1,1)$ the extremal coefficient. In the bivariate case, we can make use of the Pickands dependence function $A$ with $$\ell(u_1,u_2)=(u_1+u_2)A\left({u_2 \over u_1+u_2}\right),$$
for which Goegebeur et al.\ \cite{GGQ_IME19} provided the following estimator under censoring. With $
\tilde{X}_j := -\log F_j(X_j)$ for $j=1,2$,
one has
$$
{P}(\tilde{X}_1 > y_1, \tilde{X}_2 > y_2)
= \exp \left\{ -(y_1+y_2)A\left({y_1 \over y_1+y_2} \right)\right\}.
$$
As $\tilde{X}_t := \min \left({\tilde{X}_1/( 1-t) } ,{\tilde{X}_2 / t }\right)$ satisfies
$$
{P}(\tilde{X}_t >z)= e^{-A(t)z}, \quad  z>0,
$$
we are led to the estimation of the exponential parameter $A(t)$ based on random right-censored data $ \tilde{X}_{t,i}$ for $i=1,\ldots,n$. As in \eqref{hillc}, the maximum likelihood estimator of $A(t)$ is given by
\begin{equation}
	\hat{A}(t) = \frac{\mbox{proportion of uncensored data}}{{1 \over n}\sum_{i=1}^n \tilde{X}_{t,i} }.
\end{equation}
Note that in practice $\tilde{X}_j$, $j=1,2$ is to be estimated using the Kaplan-Meier estimator of $F_j$.
\\
Confidence bounds can be derived using the approaches in Chapter \ref{CH:INFERENCE}. In \cite{GGQ_IME19}, one can also find robust estimators under this setting.\\

\noindent
{\bf Example: Censored Liability Loss Data.} The estimate $\hat{A}$ for the (Loss, ALAE) data from Figure \ref{alae1a} is discussed and plotted in Goegebeur et al.\  \cite[Fig. 10]{GGQ_IME19}.
Goegebeur et al.\ \cite{GGQ_IME20} also considered the payment
$g(X_1,X_2)$ by the reinsurer given by
\[
g(X_1,X_2) = \begin{cases}
0, & \text{if } X_1 \leq M, \\
X_1-M + {X_1-M\over X_1}X_2,& \text{if } M < X_1 \leq M+L,\\
L + {L \over L+M}X_2, & \text{if } X_1 \geq M+L,
\end{cases}
\]
with $X_1$
representing the loss and $X_2$ the ALAE. 
The pure premium is then given by 
\[
{E}\left\{g(X_1,X_2)\right\}=\int_{[0,\infty)^2}g(x_1,x_2)\diff
\hat{F}(x_1,x_2)\]
with $\hat{F}$ following from \eqref{MEGPD2}. Numerical integration is needed here. \\

An alternative to splicing are again multivariate versions of matrix-Mittag\textendash Leffler distributions, see Albrecher et al.\  \cite{albrecher2020multivariate,albrecher2021multivariate}.  
%while a bivariate censoring solution was provided in Goegebeur et al.\ \cite{GGQ_IME19} through the estimation of the Pickands dependence function. 

\section{Natural Catastrophe Insurance and Climate Change}\label{sec7}\index{natural catastrophe insurance}
The importance of multivariate extremes becomes particularly apparent for the insurance of losses due to natural catastrophes, where marginal tails often are very heavy and at the same time correlation of claims and even of entire perils can be very strong. In such cases it is not even clear whether the risks are insurable, and there is a strong need for a profound understanding of the joint occurrences of extremes. In fact, each type of peril requires a quite different modelling approach in practice, see Hao et al.\ \cite{hao2018compound}. This may even include the need to define new spatial distance concepts. An example is a river-flow rather than Euclidian distance for locations in a river network for calibrating a max-stable process of Brown–Resnick type for joint discharges at river gauges, cf.\ Asadi et al.\ \cite{asadi2015extremes}; this model was then for instance used to quantify fluvial flood risk and its spatial diversification potential across a country, see Albrecher et al.\ \cite{albrecher2020spatial}. Other perils like storms may be more easily linked to measurable variables like maximum wind speeds, which can be modelled itself and then inform the storm loss models of (re)insurers, see e.g.\ Prettenthaler et al.\ \cite{prettenthaler2012risk}, Prahl et al.\ \cite{prahl2015comparison}, Lescourret and Robert \cite{lescourret2006extreme}, Mornet et al.\ \cite{mornet2015index}. Whereas a causal model for hazards and their induced losses can be seen as an eventual goal, the interplay of factors is typically much too complex to solely rely on these, and bottom-up models (like hydrological models for flood risk) are not designed to fully capture extremes, so that the toolkit of statistics of extremes (and in particular regression modelling with those components that are available) remains an essential ingredient in the analysis. Perils like hailstorms are even harder to model, see e.g.\ Allen et al.\ \cite{allen2015influence,allen2017extreme} and Miralles et al.\ \cite{miralles2023bayesian}. In some cases it can also be meaningful to let the dependence structure itself depend on covariates, see e.g.\ Mhalla et al. \cite{mhalla2019regression} as well as Chapter \ref{CH:REGMULTI}.\\

Another direction where insurance applications generate interesting research questions is in the context of changing climatic conditions. Non-stationary data in EVA have been studied already for some time, see e.g.\ Chavez-Demoulin and Davison \cite{chavez2005generalized}, but the recent evidence of climate change (see, e.g., Sobel et al.\ \cite{sobel2016human}, Koch et al.\ \cite{koch2021trends}, Seneviratne et al.\ \cite{seneviratne2021weather}) creates new questions and challenges, both in the analysis of frequency and severity of events. A sound modelling of extremes leading to an adequate quantification of catastrophe risk is crucial for a proper risk management of reinsurance companies, so that reinsurance coverage can still be provided for natural hazards in the face of climate change. \\

\noindent
{\bf Acknowledgement.} The authors would like to thank Alaric Mueller for help with the implementation of the numerical examples, and François Dufresne for a careful reading of the manuscript.

\bibliographystyle{abbrv}
\bibliography{bibtex_chapter29}

\begin{thebibliography}{10}

\bibitem{aban2006parameter}
I.~B. Aban, M.~M. Meerschaert, and A.~K. Panorska.
\newblock Parameter estimation for the truncated {P}areto distribution.
\newblock {\em Journal of the American Statistical Association},
  101(473):270--277, 2006.

\bibitem{akritas2003estimation}
M.~G. Akritas and I.~V. Keilegom.
\newblock Estimation of bivariate and marginal distributions with censored
  data.
\newblock {\em Journal of the Royal Statistical Society Series B: Statistical
  Methodology}, 65(2):457--471, 2003.

\bibitem{albrecher2010reinsurance}
H.~Albrecher.
\newblock Reinsurance.
\newblock {\em Encyclopedia of Quantitative Finance}, pages 1539--1543, 2010.

\bibitem{AAB_Astin21}
H.~Albrecher, J.~C. Araujo-Acuna, and J.~Beirlant.
\newblock Tempered {P}areto-type modelling using {W}eibull distributions.
\newblock {\em ASTIN Bulletin}, 51(2):509--538, 2021.

\bibitem{ABT17}
H.~Albrecher, J.~Beirlant, and J.~L. Teugels.
\newblock {\em Reinsurance: {A}ctuarial and {S}tatistical {A}spects}.
\newblock Wiley Ser. Probab. Stat. Hoboken, NJ: John Wiley \& Sons, 2017.

\bibitem{albrecher2022informed}
H.~Albrecher and M.~Bladt.
\newblock Informed censoring: The parametric combination of data and expert
  information.
\newblock {\em Statistical Planning and Inference}, 233:106171, 2024.

\bibitem{MML20}
H.~Albrecher, M.~Bladt, and M.~Bladt.
\newblock Matrix {Mittag}-{Leffler} distributions and modeling heavy-tailed
  risks.
\newblock {\em Extremes}, 23(3):425--450, 2020.

\bibitem{albrecher2020multivariate}
H.~Albrecher, M.~Bladt, and M.~Bladt.
\newblock Multivariate fractional phase--type distributions.
\newblock {\em Fractional Calculus and Applied Analysis}, 23(5):1431--1451,
  2020.

\bibitem{albrecher2021multivariate}
H.~Albrecher, M.~Bladt, and M.~Bladt.
\newblock Multivariate matrix {M}ittag--{L}effler distributions.
\newblock {\em Annals of the Institute of Statistical Mathematics},
  73:369--394, 2021.

\bibitem{albrecher2020spatial}
H.~Albrecher, D.~Kortschak, and F.~Prettenthaler.
\newblock Spatial dependence modeling of flood risk using max-stable processes:
  The example of {A}ustria.
\newblock {\em Water}, 12(6):1805, 2020.

\bibitem{allen2017extreme}
J.~T. Allen, M.~K. Tippett, Y.~Kaheil, A.~H. Sobel, C.~Lepore, S.~Nong, and
  A.~Muehlbauer.
\newblock An extreme value model for {US} hail size.
\newblock {\em Monthly Weather Review}, 145(11):4501--4519, 2017.

\bibitem{allen2015influence}
J.~T. Allen, M.~K. Tippett, and A.~H. Sobel.
\newblock Influence of the {E}l {N}i{\~n}o/{S}outhern oscillation on tornado
  and hail frequency in the {U}nited {S}tates.
\newblock {\em Nature Geoscience}, 8(4):278--283, 2015.

\bibitem{asadi2015extremes}
P.~Asadi, A.~C. Davison, and S.~Engelke.
\newblock Extremes on river networks.
\newblock {\em The Annals of Applied Statistics}, 9:2023--2050, 2015.

\bibitem{bakar2015modeling}
S.~A. Bakar, N.~A. Hamzah, M.~Maghsoudi, and S.~Nadarajah.
\newblock Modeling loss data using composite models.
\newblock {\em Insurance: Mathematics and Economics}, 61:146--154, 2015.

\bibitem{BFG16}
J.~Beirlant, I.~F. Alves, and I.~Gomes.
\newblock Tail fitting for truncated and non-truncated {P}areto-type
  distributions.
\newblock {\em Extremes}, 19:429--462, 2016.

\bibitem{beirlant2017fitting}
J.~Beirlant, I.~F. Alves, and T.~Reynkens.
\newblock Fitting tails affected by truncation.
\newblock {\em Electronic Journal of Statistics}, 11:2026--2065, 2017.

\bibitem{BBdWG2016}
J.~Beirlant, A.~Bardoutsos, T.~{de Wet}, and I.~Gijbels.
\newblock Bias reduced tail estimation for censored {P}areto type
  distributions.
\newblock {\em Statistics \& Probability Letters}, 109:78--88, 2016.

\bibitem{beirlant2004statistics}
J.~Beirlant, Y.~Goegebeur, J.~Segers, and J.~L. Teugels.
\newblock {\em Statistics of extremes: theory and applications}.
\newblock John Wiley \& Sons, 2004.

\bibitem{BGDF07}
J.~Beirlant, A.~Guillou, G.~Dierckx, and A.~Fils-Villetard.
\newblock Estimation of the extreme value index and extreme quantiles under
  random censoring.
\newblock {\em Extremes}, 10(3):151--174, 2007.

\bibitem{beirlant2018penalized}
J.~Beirlant, G.~Maribe, and A.~Verster.
\newblock Penalized bias reduction in extreme value estimation for censored
  {P}areto-type data, and long-tailed insurance applications.
\newblock {\em Insurance: Mathematics and Economics}, 78:114--122, 2018.

\bibitem{beirlant2019estimation}
J.~Beirlant, J.~Worms, and R.~Worms.
\newblock Estimation of the extreme value index in a censorship framework:
  Asymptotic and finite sample behavior.
\newblock {\em Journal of Statistical Planning and Inference}, 202:31--56,
  2019.

\bibitem{BAB_SAJ20}
M.~Bladt, H.~Albrecher, and J.~Beirlant.
\newblock Combined tail estimation using censored data and expert information.
\newblock {\em Scandinavian Actuarial Journal}, 2020(6):503--525, 2020.

\bibitem{bladt2021trimmed}
M.~Bladt, H.~Albrecher, and J.~Beirlant.
\newblock Trimmed extreme value estimators for censored heavy-tailed data.
\newblock {\em Electronic Journal of Statistics}, 15(1):3112--3136, 2021.

\bibitem{bladt2023expert}
M.~Bladt and C.~Furrer.
\newblock Expert {K}aplan--{M}eier estimation.
\newblock {\em Scandinavian Actuarial Journal}, 2024:1--27, 2024.

\bibitem{bladt2023censored}
M.~Bladt and I.~Rodionov.
\newblock Censored extreme value estimation.
\newblock {\em arXiv preprint arXiv:2312.10499}, 2023.

\bibitem{bladt2022heavy}
M.~Bladt and J.~Yslas.
\newblock Heavy-tailed phase-type distributions: a unified approach.
\newblock {\em Extremes}, 25(3):529--565, 2022.

\bibitem{bolviken2024loss}
E.~B{\o}lviken and I.~Hob{\ae}k~Haff.
\newblock Loss modeling with many-parameter distributions.
\newblock {\em Scandinavian Actuarial Journal}, pages 1--18, 2024.

\bibitem{calderin2016modeling}
E.~Calder{\'\i}n-Ojeda and C.~F. Kwok.
\newblock Modeling claims data with composite {S}toppa models.
\newblock {\em Scandinavian Actuarial Journal}, 2016(9):817--836, 2016.

\bibitem{chavez2005generalized}
V.~Chavez-Demoulin and A.~Davison.
\newblock Generalized additive modelling of sample extremes.
\newblock {\em Journal of the Royal Statistical Society Series C: Applied
  Statistics}, 54(1):207--222, 2005.

\bibitem{ciumara2006actuarial}
R.~Ciumara.
\newblock An actuarial model based on the composite {W}eibull-{P}areto
  distribution.
\newblock {\em Mathematical Reports Bucharest}, 8(4):401, 2006.

\bibitem{cooray2005modeling}
K.~Cooray and M.~M. Ananda.
\newblock Modeling actuarial data with a composite {L}ognormal-{P}areto model.
\newblock {\em Scandinavian Actuarial Journal}, 2005(5):321--334, 2005.

\bibitem{davison2015statistics}
A.~C. Davison and R.~Huser.
\newblock Statistics of extremes.
\newblock {\em Annual Review of Statistics and its Application}, 2:203--235,
  2015.

\bibitem{dempster1977maximum}
A.~P. Dempster, N.~M. Laird, and D.~B. Rubin.
\newblock Maximum likelihood from incomplete data via the {EM} algorithm.
\newblock {\em Journal of the Royal Statistical Society: Series B},
  39(1):1--22, 1977.

\bibitem{dierckx2021local}
G.~Dierckx, Y.~Goegebeur, and A.~Guillou.
\newblock Local robust estimation of {P}areto-type tails with random right
  censoring.
\newblock {\em Sankhy{\=a} A}, 83:70--108, 2021.

\bibitem{einmahl2008statistics}
J.~H. Einmahl, A.~Fils-Villetard, and A.~Guillou.
\newblock Statistics of extremes under random censoring.
\newblock {\em Bernoulli}, 14(1):207--227, 2008.

\bibitem{eling2019actual}
M.~Eling and J.~Wirfs.
\newblock What are the actual costs of cyber risk events?
\newblock {\em European Journal of Operational Research}, 272(3):1109--1119,
  2019.

\bibitem{engelke2021sparse}
S.~Engelke and J.~Ivanovs.
\newblock Sparse structures for multivariate extremes.
\newblock {\em Annual Review of Statistics and Its Application}, 8:241--270,
  2021.

\bibitem{fackler2013reinventing}
M.~Fackler.
\newblock Reinventing {P}areto: Fits for both small and large losses.
\newblock In {\em ASTIN Colloquium, Den Haag}, 2013.

\bibitem{fackler2022premium}
M.~Fackler.
\newblock Premium rating without losses: How to estimate the loss frequency of
  loss-free risks.
\newblock {\em European Actuarial Journal}, 12(1):275--316, 2022.

\bibitem{fung2023soft}
T.~C. Fung, H.~Jeong, and G.~Tzougas.
\newblock Soft splicing model: bridging the gap between composite model and
  finite mixture model.
\newblock {\em Scandinavian Actuarial Journal}, 2024:168--197, 2024.

\bibitem{ghaddab2023extreme}
S.~Ghaddab, M.~Kacem, C.~de~Peretti, and L.~Belkacem.
\newblock Extreme severity modeling using a {GLM}-{GPD} combination:
  application to an excess of loss reinsurance treaty.
\newblock {\em Empirical Economics}, 65:1105--1127, 2023.

\bibitem{GGQ_IME19}
Y.~Goegebeur, A.~Guillou, and J.~Qin.
\newblock Robust estimation of the {P}ickands dependence function under random
  right censoring.
\newblock {\em Insurance: Mathematics and Economics}, 87:101--114, 2019.

\bibitem{GGQ_IME20}
Y.~Goegebeur, A.~Guillou, and J.~Qin.
\newblock Extreme value estimation of the conditional risk premium in
  reinsurance.
\newblock {\em Insurance: Mathematics and Economics}, 96:68--80, 2021.

\bibitem{goegebeur2023conditional}
Y.~Goegebeur, A.~Guillou, and J.~Qin.
\newblock Conditional tail moment and reinsurance premium estimation under
  random right censoring.
\newblock {\em Test}, 33:230--250, 2024.

\bibitem{Guevara20}
W.~Guevara-Alarc{\'o}n, H.~Albrecher, and P.~Chowdhury.
\newblock On marine liability portfolio modeling.
\newblock {\em ASTIN Bulletin}, 50(1):61--93, 2020.

\bibitem{hao2018compound}
Z.~Hao, V.~P. Singh, and F.~Hao.
\newblock Compound extremes in hydroclimatology: a review.
\newblock {\em Water}, 10(6):718, 2018.

\bibitem{kiriliouk2019peaks}
A.~Kiriliouk, H.~Rootz{\'e}n, J.~Segers, and J.~L. Wadsworth.
\newblock Peaks over thresholds modeling with multivariate generalized {P}areto
  distributions.
\newblock {\em Technometrics}, 61(1):123--135, 2019.

\bibitem{klugman2012loss}
S.~A. Klugman, H.~H. Panjer, and G.~E. Willmot.
\newblock {\em Loss models: from data to decisions}, volume 715.
\newblock John Wiley \& Sons, 2012.

\bibitem{koch2021trends}
E.~Koch, J.~Koh, A.~C. Davison, C.~Lepore, and M.~K. Tippett.
\newblock Trends in the extremes of environments associated with severe {US}
  thunderstorms.
\newblock {\em Journal of Climate}, 34(4):1259--1272, 2021.

\bibitem{lee2012modeling}
D.~Lee, W.~K. Li, and T.~S.~T. Wong.
\newblock Modeling insurance claims via a mixture exponential model combined
  with peaks-over-threshold approach.
\newblock {\em Insurance: Mathematics and Economics}, 51(3):538--550, 2012.

\bibitem{lee2010modeling}
S.~C. Lee and X.~S. Lin.
\newblock Modeling and evaluating insurance losses via mixtures of {E}rlang
  distributions.
\newblock {\em North American Actuarial Journal}, 14(1):107--130, 2010.

\bibitem{lescourret2006extreme}
L.~Lescourret and C.~Y. Robert.
\newblock Extreme dependence of multivariate catastrophic losses.
\newblock {\em Scandinavian Actuarial Journal}, 2006(4):203--225, 2006.

\bibitem{li2023claims}
J.~Li and J.~Liu.
\newblock Claims modelling with three-component composite models.
\newblock {\em Risks}, 11(11):196, 2023.

\bibitem{Meerschaert2012parameter}
M.~M. Meerschaert, P.~Roy, and Q.~Shao.
\newblock Parameter estimation for exponentially tempered power law
  distributions.
\newblock {\em Communications in Statistics-Theory and Methods},
  41(10):1839--1856, 2012.

\bibitem{mhalla2019regression}
L.~Mhalla, M.~de~Carvalho, and V.~Chavez-Demoulin.
\newblock Regression-type models for extremal dependence.
\newblock {\em Scandinavian Journal of Statistics}, 46(4):1141--1167, 2019.

\bibitem{miralles2023bayesian}
O.~Miralles, A.~C. Davison, and T.~Schmid.
\newblock Bayesian modeling of insurance claims for hail damage.
\newblock {\em arXiv preprint arXiv:2308.04926}, 2023.

\bibitem{mornet2015index}
A.~Mornet, T.~Opitz, M.~Luzi, and S.~Loisel.
\newblock Index for predicting insurance claims from wind storms with an
  application in france.
\newblock {\em Risk Analysis}, 35(11):2029--2056, 2015.

\bibitem{nadarajah2014new}
S.~Nadarajah and S.~A. Bakar.
\newblock New composite models for the {D}anish fire insurance data.
\newblock {\em Scandinavian Actuarial Journal}, 2014(2):180--187, 2014.

\bibitem{pigeon2011composite}
M.~Pigeon and M.~Denuit.
\newblock Composite {L}ognormal--{P}areto model with random threshold.
\newblock {\em Scandinavian Actuarial Journal}, 2011(3):177--192, 2011.

\bibitem{poudyal2023finite}
C.~Poudyal and V.~Brazauskas.
\newblock Finite-sample performance of the {T}-and {W}-estimators for the
  {P}areto tail index under data truncation and censoring.
\newblock {\em Journal of Statistical Computation and Simulation},
  93(10):1601--1621, 2023.

\bibitem{prahl2015comparison}
B.~F. Prahl, D.~Rybski, O.~Burghoff, and J.~P. Kropp.
\newblock Comparison of storm damage functions and their performance.
\newblock {\em Natural Hazards and Earth System Sciences}, 15(4):769--788,
  2015.

\bibitem{prettenthaler2017flood}
F.~Prettenthaler, H.~Albrecher, P.~Asadi, and J.~K{\"o}berl.
\newblock On flood risk pooling in {E}urope.
\newblock {\em Natural hazards}, 88:1--20, 2017.

\bibitem{prettenthaler2012risk}
F.~Prettenthaler, H.~Albrecher, J.~K{\"o}berl, and D.~Kortschak.
\newblock Risk and insurability of storm damages to residential buildings in
  {A}ustria.
\newblock {\em The Geneva Papers on Risk and Insurance-Issues and Practice},
  37:340--364, 2012.

\bibitem{raschke2020alternative}
M.~Raschke.
\newblock Alternative modelling and inference methods for claim size
  distributions.
\newblock {\em Annals of Actuarial Science}, 14(1):1--19, 2020.

\bibitem{reynkens2017modelling}
T.~Reynkens, R.~Verbelen, J.~Beirlant, and K.~Antonio.
\newblock Modelling censored losses using splicing: A global fit strategy with
  mixed {E}rlang and extreme value distributions.
\newblock {\em Insurance: Mathematics and Economics}, 77:65--77, 2017.

\bibitem{rootzen2006multivariate}
H.~Rootz{\'e}n and N.~Tajvidi.
\newblock Multivariate generalized {P}areto distributions.
\newblock {\em Bernoulli}, 12(5):917--930, 2006.

\bibitem{rutikanga2021functional}
J.~U. Rutikanga and A.~Diop.
\newblock Functional kernel estimation of the conditional extreme value index
  under random right censoring.
\newblock {\em Afrika Statistika}, 16(2):2647--2688, 2021.

\bibitem{scollnik2007composite}
D.~P. Scollnik.
\newblock On composite lognormal-{P}areto models.
\newblock {\em Scandinavian Actuarial Journal}, 2007(1):20--33, 2007.

\bibitem{scollnik2012modeling}
D.~P. Scollnik and C.~Sun.
\newblock Modeling with {W}eibull-{P}areto models.
\newblock {\em North American Actuarial Journal}, 16(2):260--272, 2012.

\bibitem{seneviratne2021weather}
S.~I. Seneviratne, X.~Zhang, M.~Adnan, W.~Badi, C.~Dereczynski, A.~Di~Luca,
  S.~Ghosh, I.~Iskander, J.~Kossin, S.~Lewis, F.~Otto, I.~Pinto, M.~Satoh,
  S.~Vicente-Serrano, M.~Wehner, and B.~Zhou.
\newblock Weather and climate extreme events in a changing climate.
\newblock {\em Climate Change 2021: The Physical Science Basis. Contribution of
  Working Group I to the Sixth Assessment Report of the Intergovernmental Panel
  on Climate Change}, pages 1513--1766, 2021.

\bibitem{sobel2016human}
A.~H. Sobel, S.~J. Camargo, T.~M. Hall, C.-Y. Lee, M.~K. Tippett, and A.~A.
  Wing.
\newblock Human influence on tropical cyclone intensity.
\newblock {\em Science}, 353(6296):242--246, 2016.

\bibitem{stupfler2016}
G.~Stupfler.
\newblock Estimating the conditional extreme value index under random
  right-censoring.
\newblock {\em Journal of Multivariate Analysis}, 144:1--24, 2016.

\bibitem{stupfler2019}
G.~Stupfler.
\newblock On the study of extremes with dependent random right-censoring.
\newblock {\em Extremes}, 22(1):97--129, 2019.

\bibitem{verbelen2015fitting}
R.~Verbelen, L.~Gong, K.~Antonio, A.~Badescu, and S.~Lin.
\newblock Fitting mixtures of {E}rlangs to censored and truncated data using
  the {EM} algorithm.
\newblock {\em ASTIN Bulletin: The Journal of the IAA}, 45(3):729--758, 2015.

\bibitem{wang2019focussed}
Y.~Wang and I.~Hob{\ae}k~Haff.
\newblock Focussed selection of the claim severity distribution.
\newblock {\em Scandinavian Actuarial Journal}, 2019(2):129--142, 2019.

\bibitem{wang2020modelling}
Y.~Wang, I.~Hob{\ae}k~Haff, and A.~Huseby.
\newblock Modelling extreme claims via composite models and threshold selection
  methods.
\newblock {\em Insurance: Mathematics and Economics}, 91:257--268, 2020.

\bibitem{weissman1978}
I.~Weissman.
\newblock Estimation of parameters and large quantiles based on the $k$ largest
  observations.
\newblock {\em Journal of the American Statistical Association},
  73(364):812--815, 1978.

\bibitem{WW2016}
J.~Worms and R.~Worms.
\newblock New estimators of the extreme value index under random right
  censoring, for heavy tailed distributions.
\newblock {\em Extremes}, 17:337--358, 2014.

\end{thebibliography}

\printindex
%\cleardoublepage

\end{document}